%% file: main.tex
\documentclass[fleqn,usenatbib]{mnras}

\usepackage{newtxtext,newtxmath}
\usepackage[T1]{fontenc}

\DeclareRobustCommand{\VAN}[3]{#2}
\let\VANthebibliography\thebibliography
\def\thebibliography{\DeclareRobustCommand{\VAN}[3]{##3}\VANthebibliography}

\usepackage{graphicx}
\usepackage{amsmath}
\usepackage{multirow}
\usepackage{arydshln}
\usepackage{bm}
\usepackage[normalem]{ulem}

\usepackage{hyperref}

\title[Converting the sub-Jovian desert to a savanna]{Converting the sub-Jovian desert of exoplanets to a savanna with TESS, PLATO and Ariel}
\author[Sz. K\'alm\'an et al.]{
Szil\'ard K\'alm\'an,$^{1,2,3,4,5}$\thanks{E-mail: xilard1@gothard.hu}
Gyula M. Szab\'o,$^{2,6,7}$
Luca Borsato,$^{8}$
Attila B\'odi,$^{1,5,10}$
Andr\'as P\'al$^{1,5}$\newauthor
and R\'obert Szab\'o$^{1,5,9,10}$
\\
$^{1}$Konkoly Observatory, Research Centre for Astronomy and Earth Sciences, ELKH,  MTA CSFK Lend\"ulet Near-Field Cosmology Research Group, \\Konkoly-Thege Miklós út 15–17., H-1121, Hungary \\
$^{2}$MTA-ELTE Exoplanet Research Group, Szombathely, Szent Imre h. u. 112., H-9700, Hungary \\
$^{3}$ELTE E{\"o}tv{\"o}s Lor\'and University, Doctoral School of Physics,  Budapest, Pázmány Péter sétány 1/A, H-1117, Hungary \\
$^{4}$Graduate School of Physics, University of Szeged, Szeged, D\'om t\'er 9., H-6720, Hungary \\
$^{5}$CSFK, MTA Centre of Excellence, Budapest, Konkoly Thege Miklós út 15-17., H-1121, Hungary \\
$^{6}$ELTE E{\"o}tv{\"o}s Lor\'and University, Gothard Astrophysical Observatory, Szombathely, Szent Imre h. u. 112., H-9700, Hungary \\
$^{7}$MTA-ELTE  Lend{\"u}let "Momentum" Milky Way Research Group, Hungary\\
$^{8}$INAF-Osservatorio Astronomico di Padova a Vicolo dell’Osservatorio 5, 35122, Padova, Italy \\
$^{9}$ E\"otv\"os Lor\'and University, Institute of Physics, P\'azm\'any P\'eter s\'et\'any 1/A, H-1171 Budapest, Hungary
$^{10}$MTA CSFK Lend\"ulet Near-Field Cosmology Research Group
}

\date{Accepted XXX. Received YYY; in original form ZZZ}
\pubyear{2022}
\begin{document}
\label{firstpage}
\pagerange{\pageref{firstpage}--\pageref{lastpage}}
\maketitle

\begin{abstract}
     There is a lack of exoplanets with sizes similar to Neptune orbiting their host stars with periods $\lesssim 3$ days -- hence the name ``sub-Jovian/Neptune desert''. Recently, several exoplanets have been confirmed to reside in the desert transforming it into a ``savanna'' with several ``giraffe'' planets (such as LTT 9779 b and TOI-674 b).
    The most prominent scenarios put forward for the explanation of the formation of the desert are related to the stellar irradiation destroying the primary atmosphere of certain specific exoplanets. We aim to present three targets (LTT 9779 b, TOI-674 b and WASP-156 b) which, when observed at wide wavelength ranges in infrared (IR), could prove the presence of these processes, and therefore improve the theories of planetary formation/evolution.
    We simulate and analyse realistic light curves of the selected exoplanets with PLATO/NCAM and the three narrow-band filters of Ariel (VISPhot, FGS1 and FGS2) based on TESS observations of these targets.
    We improved the precision of the transit parameters of the three considered planets from the TESS data. We find that the combination of the three narrow-band filters of Ariel can yield inner precision of $\lesssim 1.1\%$ for the planetary radii. Data from the three telescopes together will span decades, allowing the monitoring of changes in the planetary atmosphere through radius measurements.
    The three selected ``giraffe'' planets can be golden targets for Ariel, whereby the loss of planetary mass due to stellar irradiation could be studied with high precision, multi-wavelength (spectro-)photometry.
\end{abstract}


\section{Introduction}

An enigmatic ``hole'' in the exoplanet distribution, the lack of exoplanets ranging between Jupiters and super-Earths in size and with orbital periods $\lesssim 3$ days was reported by \cite{2011ApJ...727L..44S} and was later confirmed by \cite{2016A&A...589A..75M}. This region in the mass--period ($M_P$ -- $P$) and radius--period ($R_P$ -- $P$) parameter spaces was named the sub-Jovian/Neptune desert of exoplanets. The presence of an ever-increasing number of objects in the sub-Jovian desert such as NGTS-4b \citep{west}, NGTS-5b \citep{2019A&A...625A.142E},  LTT 9779b \cite{2020NatAs...4.1148J}, TOI-849b \citep{armstrong}, TOI-674b \citep{murgas}, NGTS-14Ab \citep{2021A&A...646A.183S}, or TOI-2196b \citep{2022arXiv220805797P} suggests that it is not completely empty. With the increasing number of confirmed exoplanets, it is reasonable to expect that the number of known exoplanets will increase in the future. Because TOI-674 b resides inside the sub-Jovian desert, \cite{murgas} introduced the term ``oasis'' for it. The detection of water vapor in its atmosphere \citep{brande} could also be in line with this terminology. However, without the confirmation that all planets in this area of the parameter spaces have a water-rich atmosphere, we suggest the term ``sub-Jovian savanna'' for the area in question (see \cite{2023arXiv230101065S} for details), while calling the individual planets ``giraffes'' in the savanna. The naming of the objects is justified as these planets were the first (i.e. the easiest) to be discovered in the savanna, similarly to their namesake animals, which can easily be seen from a long distance.



There are a number of proposed explanations for the savanna (see \cite{2018MNRAS.479.5012O} and \cite{2023arXiv230101065S} for an overview), however, photoevaportaion of close-in exoplanets seems to play an essential role in its formation (\cite{2018MNRAS.479.5012O, 2019MNRAS.485L.116S}. In \cite{2023arXiv230101065S}, we explored the dependencies of the savanna across the HRD parameters in search of correlations and implications on how these regions of the period-mass and period-radius parameter spaces are formed, confirming the role of photoevaporation while also pointing out that it alone cannot explain this phenomenon. In this paper, we focus on the observational aspects of the understanding of the savanna. For this purpose, we selected three known exoplanets, LTT 9779 b, TOI-674 b, and WASP-156 b \citep{demangeon} which are in the savanna. 

The \verb|TESS| \citep{ricker} light curve from S2 of LTT 9779 b was analyzed by \cite{2020NatAs...4.1148J}. \cite{crossfield} and \cite{dragomir} have explored its bulk composition and atmosphere based on \verb|Spitzer| IRAC \citep{fazio} observations. Ground-based transit observations from the Las Cumbres Observatory and the Next Generation Transit Survey \citep{ngts} were also included in the analysis that confirmed the planetary nature of LTT 9779 b. The planetary nature of the companion of the nearby M dwarf TOI-674
was confirmed by \cite{murgas} with the help of ground-based and \verb|Spitzer| observations. Both \cite{murgas} and \cite{brande} reported that the super-Neptune is present in the Neptune desert. In their analysis, \cite{brande} analyzed the \verb|TESS| light curves from S9, S10 and S36, while also incorporating data obtained by the Hubble Space Telescope with a focus on the planetary atmosphere. The super-Neptune WASP-156 b was discovered by \cite{demangeon}. \cite{2021AJ....162..221S} analyzed the \verb|TESS| light curves from S4 and S31 via wavelet denoising and GP regression techniques. The transit parameters were also derived by \cite{yangfan} based on \verb|TESS| observations and the Gaia database \citep{gaia16, gaia18}.


In this paper, we aim to demonstrate how direct observational evidence can be gained by observing the giraffe exoplanets for testing the atmospheric mass-loss-based  scenarios \citep{2023arXiv230101065S} for the evolution of the savanna via observations with \verb|PLATO| \citep[Planetary Transits and Oscillations of stars]{2014ExA....38..249R} and \verb|Ariel| \citep[Atmospheric remote-sensing infrared exoplanet large survey;][]{2021arXiv210404824T}. We propose that by constraining the transit depth through optical and near-infrared (NIR) measurements, the wavelength-dependence of the transit depth, expected in IR, can be attributed to the loss of an extended atmosphere of the giraffes. Such observations will be possible with Ariel.

This paper is structured as follows. In Sec. \ref{sec:meth}, we discuss the \verb|TESS| observations of the three selected planets, present our methods for the derivation of the transit parameters, and show the simulated light curves of \verb|Ariel| and \verb|PLATO|. In Sec. \ref{sec:tess}, the improved transit parameters based on the \verb|TESS| light curves are presented, while we show the results from the simulated light curves in Sec. \ref{sec:arielplato}. Finally, we discuss the time- and wavelength-dependence of $R_P$ and the observational prospects presented by the upcoming ESA missions (Sec. \ref{sec:disco}).


\section{Methods} \label{sec:meth}

In this section, we describe the light curve solutions for the available \verb|TESS| data, as well as the simulations for the Ariel and \verb|PLATO| missions.

\begin{figure}
    \centering
    \includegraphics[width=0.48\textwidth]{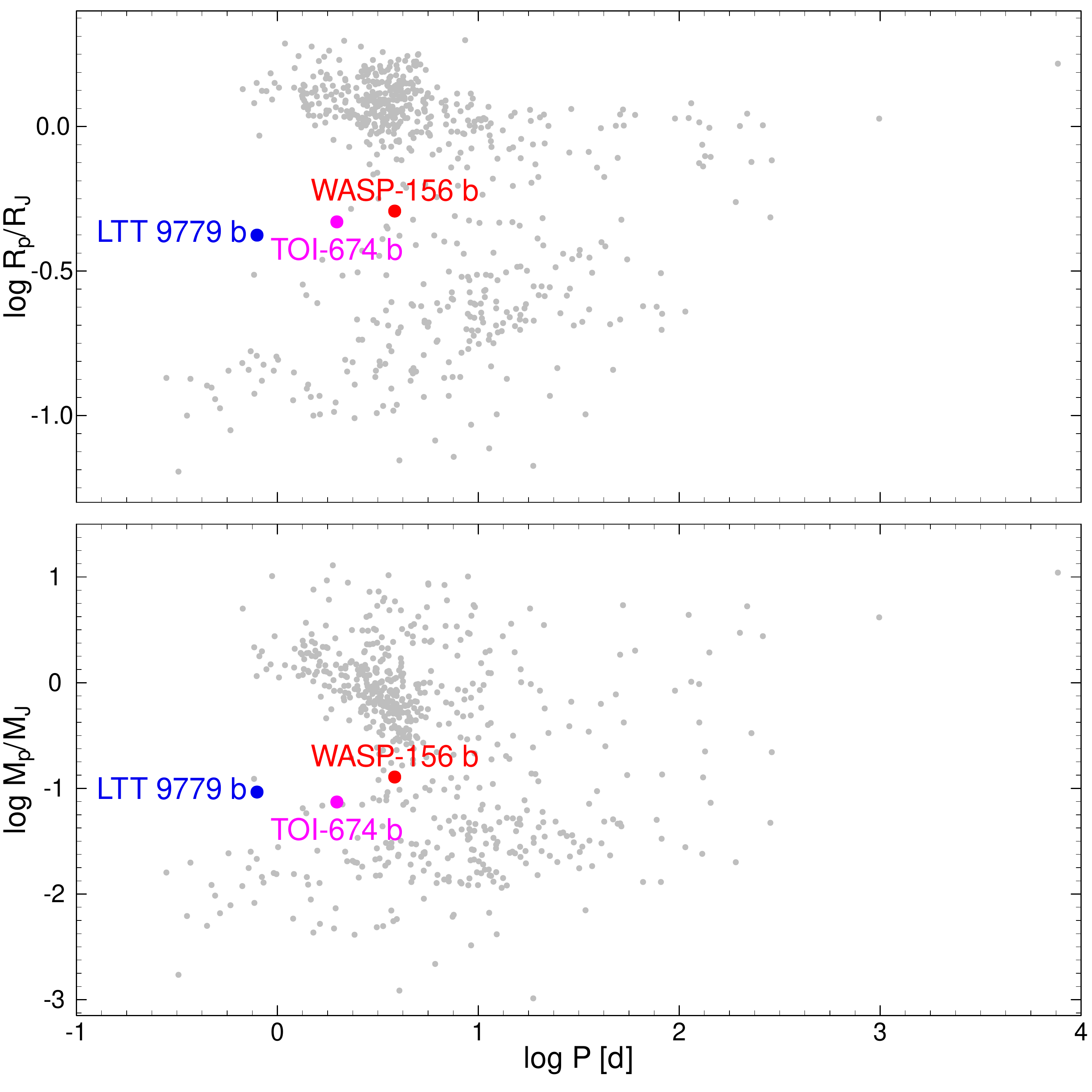}
    \caption{Position of the three selected planets in the radius-period and mass-period planes. LTT 9779 b and TOI-674 b clearly populate the sub-Jovian savanna in both cases, while WASP-156 b is on the border. The planetary parameter database is the same as in \protect\cite{2023arXiv230101065S}.}
    \label{fig:giraffes}
\end{figure}

\subsection{Target selection: populating the desert}

From the sample of all known exoplanets in the NASA Exoplanet Archive\footnote{\url{https://exoplanetarchive.ipac.caltech.edu/}}, we selected three planets (LTT 9779 b, TOI-674 b and WASP-156 b, Fig. \ref{fig:giraffes}) the detailed analysis of which can lead to obtaining observational evidence of how the savanna is formed. 
We selected LTT 9779 as it clearly resides in the savanna in both $M_P$ -- $P$ and $R_P$ -- $P$ parameter planes (Fig. \ref{fig:giraffes}). Given that there is a small number of known planets in these regions of the parameter planes, the other two objects so that they are on the boundaries: TOI-674 b is ``inside'' only in the $R_P$-- $P$ plane (upper panel of Fig. \ref{fig:giraffes}), while WASP-156 b is on the border in both parameter spaces. A further criteria for the selection was the availability of TESS observations.
We focus on the information that can be gained from the best currently available optical light curves and show the validity of future observations with the next generation of space-based observatories.

\subsection{Observations with TESS}

All three systems were observed by \verb|TESS| in at least two sectors: S2 and S29 for LTT 9779; S9, S10 and S36 for TOI-674; S3, S31, S42 and S43 for WASP-156. We used the \verb|lightkurve| software package  \citep{lightkurve} to obtain the 2-min cadence Simple Aperture Photometry (SAP) light curves of all three targets. 
We removed all measurements with non-zero quality flags and combined the data from all sectors of each star, which resulted in light curves of $34232$, $46427$ and $58041$ data points for LTT 9779, TOI-674 and WASP-156, respectively.  In the case of LTT 9779, we also removed data containing the first transit because of severe baseline variations. We also removed several artefacts from the SAP flux light curve of WASP-156 around 2490 BTJD.

\subsection{Light curve solutions}

We made use of the Transit and Light Curve Modeller  \citep[TLCM;][]{tlcm} code to analyze the light curves. This software incorporates the analytic Mandel-Agol transit model \citep{mandelagol} with four parameters: the scaled semi-major axis, $a/R_S$, the ratio of the planetary and stellar radii, $R_P/R_S$, the impact parameter, $b\!=\!a/R_S\cos i$ (where $i$ is the orbital inclination relative to the line of sight) and the time of midtransit, $t_C$. These parameters, as well as the orbital period $P$, were left free in all three cases. We made use of \verb|PyLDTk| \citep{ldtk} to compute the quadratic limb-darkening coefficients $u_1$ and $u_2$. The PHOENIX-generated stellar spectra used for these calculations \citep{Husser2013} were selected based on $T_{\rm eff}$, $\log g$ and [Fe/H] (Table \ref{tab:stellar}). During the fitting process, TLCM uses $u_+ = u_1+u_2$ and $u_- = u_1-u_2$.

TLCM also relies on the wavelet-formulation of \cite{carterwinn} to handle time-correlated noise, which arises as the combination of stellar variability and instrumental effects. In this approach, the noise is described by two parameters: $\sigma_r$ for the red component and $\sigma_w$ for the white component. The fitting process is then done for the noise plus transit model simultaneously. The wavelet-based noise handling approach of TLCM has a regularization condition to avoid overfitting the noise: the standard deviation of the residuals are decreased until they are equal to the average photometric uncertainty of the input light curve \citep{pow1}. While $\sigma_w$ is representative of the amount of white noise present in the light curves, $\sigma_r$ is not related to the root mean squared of the red noise \citep{carterwinn, pow1}. The usage of the wavelet-formulation was tested on noise sources with timescales ranging from several tens of seconds \citep{2022arXiv220801716K} to tens of minutes \citep{wasp33} to several hours or days \citep{pow1}, and it was found to yield consistent results in every case.
We assumed circular orbits for all three planets and the adopted stellar parameters are shown in Table \ref{tab:stellar}.

\subsection{Simulated Ariel observations} \label{sec:arielsim}

\begin{figure}
    \centering
    \includegraphics[width = .48\textwidth]{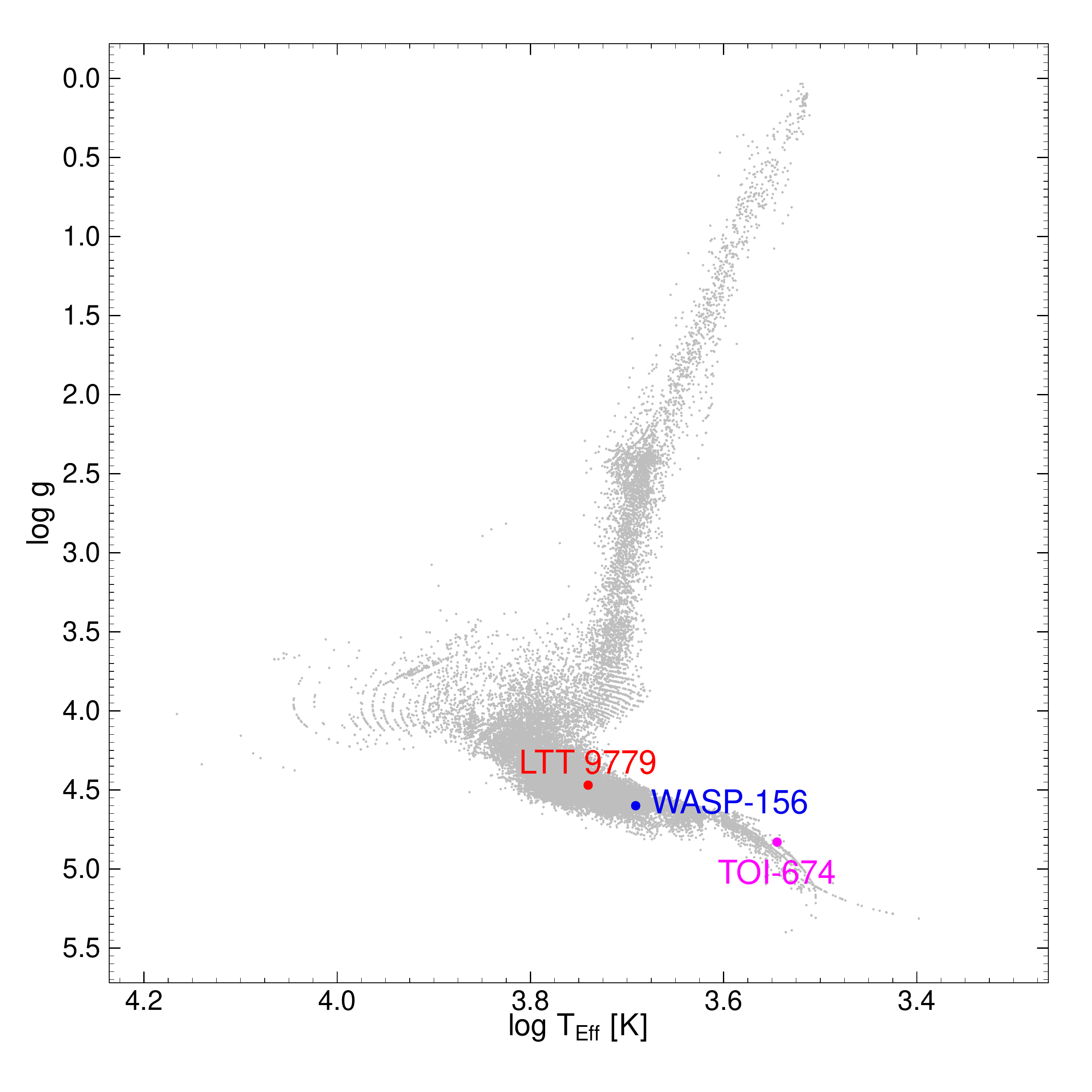}
    \caption{Kiel diagram showing the stars from \citet{huber2014}. The position of the three host stars was used for selecting relevant Kepler light curves for the simulated Ariel observations (see text for details).}
    \label{fig:hrd}
\end{figure}

Our simulations were focused on the three narrow band filters that will be available for Ariel observations \citep{2021arXiv210404824T, szabo2021}: VISPhot, FGS1 and FGS2. At the time of writing, there was no available simulator dedicated for Ariel observations. Neglecting out-of-transit variations, the observed light curves of exoplanet-hosting stars consist of three parts: transit signal; white noise and red noise. The white component was assumed to be made up of four sources \citep{arielrad}: ($i$) Poisson noise term, ($ii$) detector noise term, ($iii$) jitter noise term and ($iv$) the Payload noise floor. The noise budget for this component was calculated by ArielRad \citep{arielrad} and it is different at the three used wavelengths. The time-correlated noise (see e.g. Figs. \ref{fig:ltt9779lc}, \ref{fig:toi674full} and \ref{fig:wasp156lc}) can arise from several sources, including instrumental effects and astrophysical phenomena (including flares, granulation, pulsation, etc.) and is therefore dependent on both stellar and instrumental parameters. The exact nature of correlated noise expected from the Ariel detectors is not known. We therefore proceeded to use \verb|Kepler| observations of similar stars to those which host the three giraffe planets. These targets were identified from the database of \cite{huber2014} based on the stellar parameters from Table \ref{tab:stellar}, the position of LTT 9779, TOI-674 and WASP-156 on the Kiel diagram from this database is shown in Fig. \ref{fig:hrd}. After downloading two adjacent Quarters of long cadance SAP flux light curves for each target via the \verb|lightkurve| software package, we detrended and fitted an Autoregressive Integrated Moving Average (ARIMA) model to each of them.
The noise levels for the two noise types (white and red) of the three stars observed by Kepler are listed in Table \ref{tab:noise}. These quantities are calculated as the standard deviation of the residuals of the ARIMA processes and the fitted ARIMA processes themselves in a one-hour bin. Note that during the selection of the Kepler targets, we did not take brightness into account, resulting in higher white noise levels in comparison with Ariel.


The color-dependence of the correlated noise is not well understood. Based on prior experience, it is clear that the amplitude of the red noise decreases with longer wavelength, but no exact relationship has been established yet. Because Ariel will be able to perform simultaneous observations with different filters, it will also provide opportunities to handle this question.We simulated three separate ARIMA models with the Kepler cadence for all three stars. Then we normalized the standard deviations of these red noise models. Finally, we scaled the noise levels in the three filters by heuristically assuming that noise amplitude in VISPhot will be the same as in the Kepler observations, while in FGS1 and FGS2, the amplitude was set to $0.95 \times$ and $0.90\times$ the Kepler-levels.
We then proceeded to interpolate the 1800 s cadence noise models to $1$ s, the lowest nominal exposure time of Ariel.The red noise levels listed in Table \ref{tab:noise} are calculated as the standard deviation of these time series, binned to one hour intervals. Note that the time-correlated noise can not be interpreted as Gaussian, therefore, the variance of the binned data does not scale with the length of the bins. This is the reason behind the apparent discrepancy between the Kepler and Ariel noise levels seen in Table \ref{tab:noise}. Finally, we added a Gaussian noise model with standard deviation derived from the noise budget of ArielRad, yielding a conservative total noise estimate. The white noise levels, computed as the standard deviation of the data binned to one-hour intervals, are also listed in Table \ref{tab:noise}.


\input{noise_table.tex}

For the synthetic transits, we made use of the \verb|batman| software package \citep{batman}, which incorporates a Mandel-Agol model. Using PyLDTk, based on the stellar parameters of Table \ref{tab:stellar}, we calculated the four parameter limb-darkening coefficients \citep{claret2000} for all three stars (Table \ref{tab:ldcs}), and created the synthetic transit light curves from our derived transit parameters of Tables \ref{tab:ltt9779}, \ref{tab:toi674} and \ref{tab:wasp156} at the $1$ s cadence. 

After injecting the transits (with fixed $P$ and $t_C$) into the noise model, we binned the light curves to $60$ seconds for LTT 9779 b and TOI-674 b and to $30$ seconds for WASP-156 b to mimic the real observations that will be made with Ariel. This also allowed us to have a realistic estimate of the photometric uncertainties, which are used by TLCM to constrain the correlated noise (see \cite{pow1} for more details). 

We note that the observing strategy of Ariel will be completely different than in the case of Kepler, therefore, the systematic noise sources can be expected to differ significantly as well. So far, however, no specific instrumental noise models exist for Ariel. Therefore, our assumption of a Kepler-like red noise profile that is scaled according to wavelength is not perfectly representative of the so far unknown Ariel systematics, however, it is still considerably more realistic than assuming white noise alone.

\subsection{Simulated PLATO observations}

\verb|PLATO| Solar-like Light-curve Simulator (PSLS; \citealt{psls}) is a tool for generating realistic light curves representative of the performace of the \verb|PLATO| mission. \texttt{PSLS} is designed to perform hare-and-hound exercises to characterize the expected performance of the multi-camera concept. This tool can simultaneously simulate solar-like oscillations, stellar granulation, and magnetic activity along with random noise and systematic errors. The latter two is based on pixel level simulations and contain the effect of periodic mask updates, introducing jumps in the light curves.

In this paper, we used the publicly available \texttt{PSLS}\footnote{https://sites.lesia.obspm.fr/psls/} (v1.4) python software. To simulate light curves we only considered instrumental noise and systematic effects. It is important to note that contrary to the simulated \verb|Ariel| observations, all sources of stellar noise have been omitted. We simulated 90 days long observation sequences, where the random noise was taken from tables generated from realistic light curves. We considered two cases: in the most favorable case, the giraffes may be observed by all four camera groups (each consisting of six cameras), while in the less favorable case, they may be observed with only one group (see e.g. \cite{nascimbeni} and \cite{heller22} for details). When the ideal circumstances are present (i.e. observations with all $24$ so-called normal cameras are feasible, we averaged the data of all four camera groups with all six cameras each to obtain the final light curves, while in the other case we used the \texttt{PSLS} data for one group. For long-term stellar drift, we simulated the worst-case scenario, setting the drift level to its maximum value. The input stellar magnitudes and effective temperatures were set according to the given target's values. All the other \texttt{PSLS} parameters were kept at their default values.

To account for the possible stellar activity, we also included the noise models from the Kepler observations as described in Sect. \ref{sec:arielsim}. This way, we also introduce further systematic effects that were characteristic of the Kepler observations, however, we get more conservative estimates for the retrieved parameters. The added red noise was scaled as in the case of VISPhot. The noise levels, calculated as in Sect. \ref{sec:arielsim}, are also listed in Table \ref{tab:noise} for the two considered cases: observations with $6$ and $24$ cameras. We then injected the planetary signals in the same way as for the \verb|Ariel| simulations: the light curves were created with the \texttt{batman} software with transit parameters derived from the \verb|TESS| light curves, while using the 4-parameter limb-darkening coefficients from Table \ref{tab:ldcs}. PLATO limb-darkening coefficients were calculated using \texttt{ExoTETHyS} \citep{2020AJ....159...75M, 2020JOSS....5.1834M}. We employed state-of-the-art spectral response of the PLATO
instruments, the MARCS stellar atmosphere grid  \citep{2008A&A...486..951G}
and Non-LTE Turbospectrum models \citep{2022arXiv220600967G} computed within
the PLATO consortium. The photometric uncertainty was set to the standard deviation of the noise model. We used the nominal cadence of $25$ s (with $21$ s exposure times).

It is important to point out that we used \texttt{TLCM} to solve the light curves in every case, which also yielded consistent uncertainty estimations, which enabled straight-forward comparisons of the transit parameters between the different telescopes/instruments. We chose $30$-day-long simulated light curves for both the \verb|Ariel| and \verb|PLATO| analyses, because a time series of this length contains roughly the same number of transits as a single Sector of \verb|TESS|.

\begin{table*}
    \centering
        \caption{Limb-darkening coefficients of the three host stars for PLATO and the three narrow-band filters of Ariel.}
    \label{tab:ldcs}
    \begin{tabular}{l l r r r r | r r}
    \hline
    \hline
      \multirow{2}{*}{Host name} & \multirow{2}{*}{Instrument}  & \multicolumn{4}{c}{4-parameter law} & \multicolumn{2}{c}{Quadratic law} \\
     & & $u_1$ & $u_2$ & $u_3$ & $u_4$ & $u_1$ & $u_2$ \\
        \hline
      \multirow{4}{*}{LTT 9779} & Ariel/VISPhot & $0.084$ & $0.101$ & $1.074$& $-0.518$& $0.725$ & $0.059$\\
       & Ariel/FGS1 & $-0.128$ & $1.031$& $-0.227$& $-0.041$& $0.544$ & $0.110$\\
       & Aiel/FGS2 & $-0.139$ & $1.133$ & $-0.555$& $0.090$& $0.410$ & $0.120$\\
       & PLATO/NCAM & $0.632$ & $-0.345$& $0.807$& $-0.325$& $0.517$ & $0.134$\\
       \hline
        \multirow{4}{*}{TOI-674} & Ariel/VISPhot & $-0.142$ &$2.015$ & $-1.501$ & $0.438$ & $0.552$ & $0.231$ \\
        & Ariel/FGS1 & $0.065$ & $1.634$& $-1.459$ & $0.461$ & $0.380$ & $0.245$ \\
        & Aiel/FGS2 & $0.687$ & $-0.128$ & $0.133$ & $-0.084$ & $0.235$ & $0.210$ \\
        & PLATO/NCAM & $0.729$ & $0.298$ & $-0.226$ & $0.061$ & $0.389$ & $0.291$ \\
        \hline
        \multirow{4}{*}{WASP-156} & Ariel/VISPhot & $0.063$ & $0.524$ & $0.356$& $-0.167$ & $0.754$ & $0.021$ \\
        & Ariel/FGS1 & $0.031$& $0.842$ & $-0.194$& $0.001$ & $0.565$ & $0.099$ \\
        & Aiel/FGS2 &$0.033$ & $0.888$ & $-0.439$ & $0.092$ & $0.424$ & $0.115$ \\
        & PLATO/NCAM & $0.566$ & $-0.311$ & $0.865$ & $-0.341$ & $0.580$ & $0.099$ \\
        \hline
    \end{tabular}
\end{table*}

\section{Transit parameters from TESS} \label{sec:tess}

Following the recipe of Sec. \ref{sec:meth}, we made use of TLCM to derive transit parameters from the \verb|TESS| light curves, which were subsequently used in the Ariel and PLATO simulations.
\subsection{LTT 9779 b}

\begin{figure*}
    \centering
   \includegraphics[width = \textwidth]{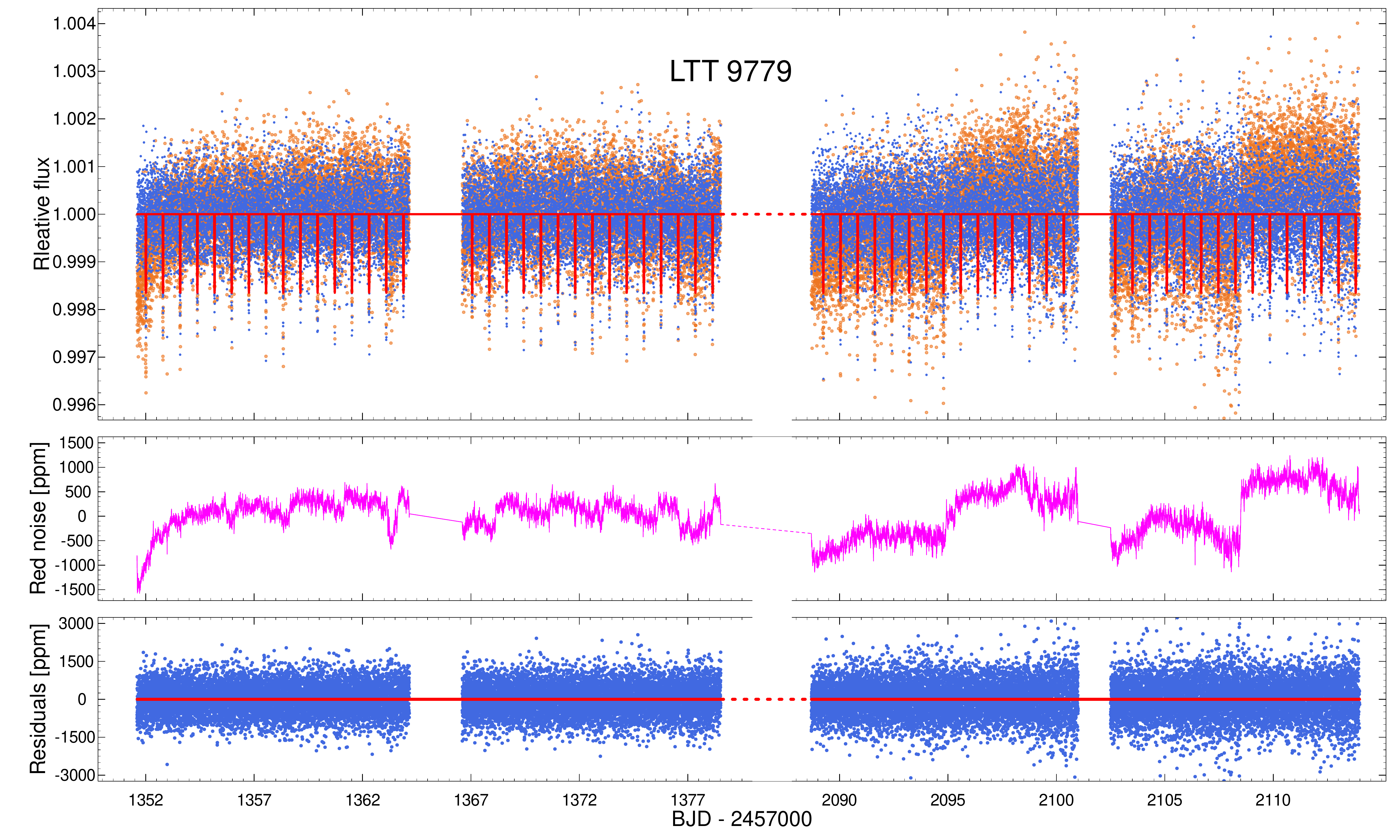}
       \caption{Observed TESS light curve of LTT 9779 from Sectors 2 and 29 (top panel, orange dots) -- note the break on the time axis. Subtracting the fitted red noise (middle panel) from the observations yields the blue dots of the top panel. The best fit transit model is shown with continuous red line. The residuals are plotted on the bottom panel.}
    \label{fig:ltt9779lc}
\end{figure*}

Our analysis of LTT 9779 b is based on two sectors of SAP flux data (Fig. \ref{fig:ltt9779lc}, top panel). We adopted the following stellar parameters for the host from \cite{2020NatAs...4.1148J}: $T_{eff} = 5499 \pm 50$ K, $\log g = 4.47 \pm 0.10$, $[Fe/H] = 0.31 \pm 0.08$ and $R_S = 0.92 \pm 0.01\ R_\odot$ (Table \ref{tab:stellar}). 
In the cases of the light curves of the giraffes, the S/N was not high enough to fit the limb-darkening parameters, even when applying Gaussian priors to them, so we opted to fix them at their theoretical values as calculated by PyLDTk ($u_{+, \text{TESS}} = 0.62$ and $u_{-, \text{TESS}} = 0.39$ for LTT 9779).

\begin{figure}
    \centering
    \includegraphics[width = 0.48\textwidth]{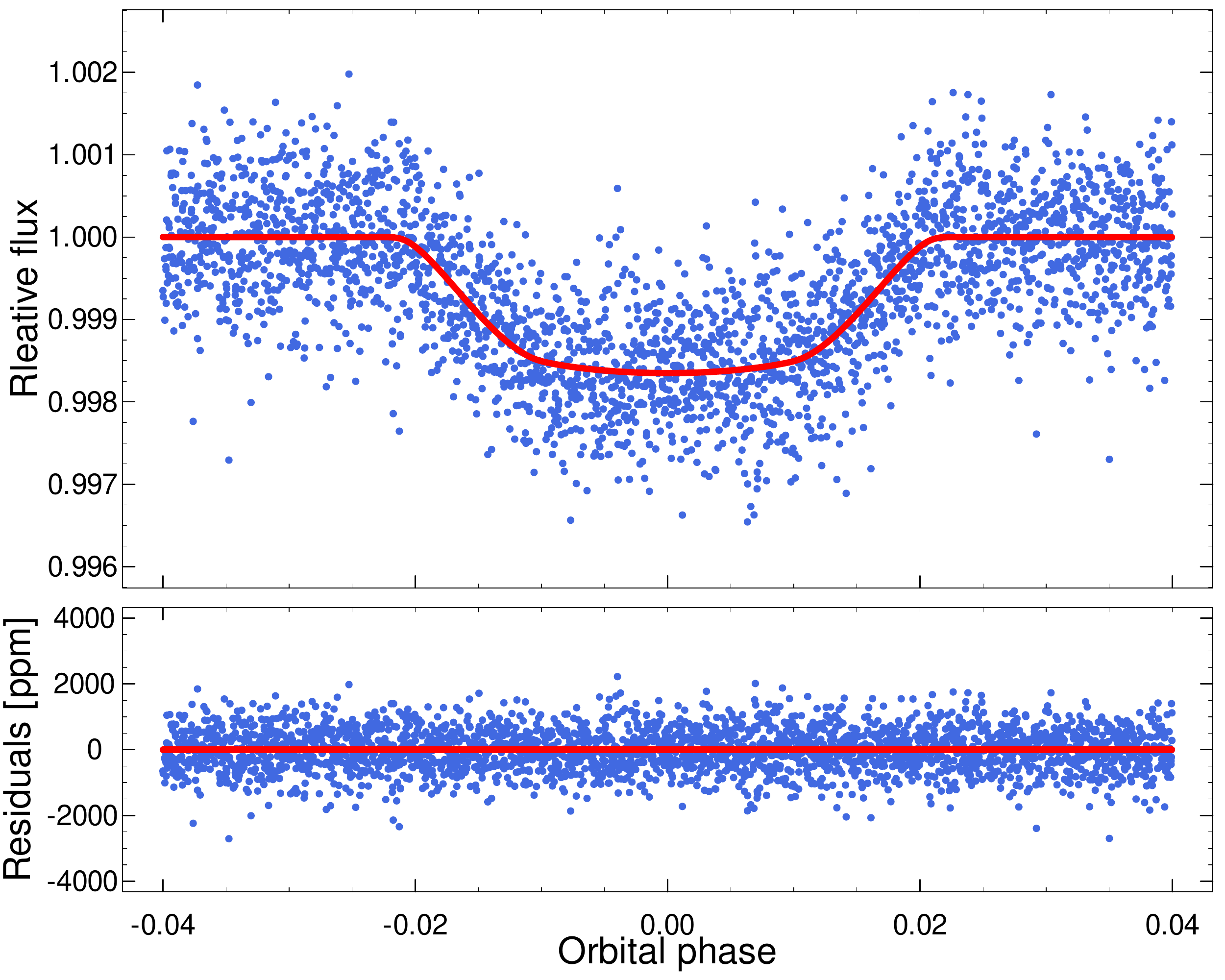}
    \caption{Phase-folded TESS light curve of LTT 9779 without correlated noise (top panel, blue dots), the best fit transit model (top panel, solid red line) and the residuals (bottom panel).}
    \label{fig:ltt9779ph}
\end{figure}

The best-fit transit model is shown on the upper panel of Fig. \ref{fig:ltt9779lc} (as a function of time) and Fig. \ref{fig:ltt9779ph} (as a function of orbital phase). The residuals of the best-fit models (lower panels of Figs. \ref{fig:ltt9779lc} and \ref{fig:ltt9779ph}) indicate that much of the correlated noise (middle panel of \ref{fig:ltt9779lc}) was removed as a consequence of the wavelet-based noise handling. The fitted transit parameters are shown in Table \ref{tab:ltt9779}. These parameters are in excellent agreement with the results of \cite{2020NatAs...4.1148J} (except for the noise terms of TLCM). Our analysis also improved the precision of the transit parameters which is the result of the correlated noise treatment of TLCM and basically double the available photometric data. The greatest improvements are observable at the orbital period with $P  =   0.79206447 \pm 2.9 \cdot 10^{-7}$ days (an uncertainty of $0.025$ seconds) and the ratio of the planetary and stellar radii with $R_P/ R_S = 0.04578 \pm 0.00066$. The latter translates to an absolute radius of $R_P = 0.419 \pm 0.061\ R_J$, locating the planet firmly in the sub-Jovian savanna, and qualifying it as a giraffe.

The white noise parameter, $\sigma_w$, roughly corresponding to the standard deviation of the residuals, implies $90$ ppm $\sqrt{\rm h}$ noise level in a one-hour bin, which is almost double of the noise level expected in FGS2 (Table \ref{tab:noise}).

\begin{table}
\caption{Stellar parameters used in the light curve analysis. Adopted from \citet{2020NatAs...4.1148J} (LTT 9779), \citet{murgas} (TOI-674) and \citet{demangeon} (WASP-156).}
\label{tab:stellar}
    \centering
    \begin{tabular}{l c c c}
    \hline
    \hline
   Parameter  & LTT 9779 &  TOI-674 &  WASP-156  \\
    \hline
    $T_{\rm eff}$ [K] &$5499 \pm 50$ & $3505 \pm 30$ & $4910 \pm 61$ \\
    $\log g$ & $4.47 \pm 0.10$ & $4.83 \pm 0.01$ & $4.60 \pm 0.05$\\
    $[Fe/H]$ & $0.31 \pm 0.08$ & $0.114 \pm 0.074$ & $0.24 \pm 0.12$\\
    $R_S / R_{\odot}$ & $0.92 \pm 0.01$ & $0.42 \pm 0.01$ & $0.76 \pm 0.04$\\
    $V$ [mag] & $9.76$ & $14.203$ & $11.59$ \\
    \hline
    \end{tabular}
\end{table}

\begin{table}
\caption{Comparison of our derived transit parameters for LTT 9779b to the results of \citet{2020NatAs...4.1148J}. The time of midtransit is expressed in BJD - 2457000.}
\label{tab:ltt9779}
\centering
\footnotesize
\begin{tabular}{l c c }
\hline
\hline
Parameter & This work & \cite{2020NatAs...4.1148J}\\
\hline
\vspace{0.1cm}
\hspace{-0.2cm} $a / R_S$ & \hspace{-0.3 cm} $3.777 \pm 0.053$ & \hspace{-0.45cm}$3.877^{+0.090}_{-0.091}$ \\ 
\vspace{0.1cm}
\hspace{-0.2cm} $R_P / R_S$ & \hspace{-0.3 cm} $0.04578 \pm 0.00066$ & \hspace{-0.45cm}$0.0455^{+0.0022}_{-0.0017}$\\ 
\vspace{0.1cm}
\hspace{-0.2cm} $b$ & \hspace{-0.3 cm} $0.9241 \pm 0.0025$ &  \hspace{-0.45cm}$0.912_{-0.049}^{+0.050}$\\ 
\vspace{0.1cm}
\hspace{-0.2cm} $P$ [days] & \hspace{-0.3 cm} $0.79206447 \pm 2.9 \cdot 10^{-7}$ & \hspace{-0.45cm}$0.7920520 \pm 9.3 \cdot 10^{-7}$\\ 
\vspace{0.1cm}
\hspace{-0.2cm} $t_C$ [TBJD] & \hspace{-0.3 cm} $1355.00699 \pm 0.00018$ & \hspace{-0.45cm} $1354.21430 \pm 0.00025$\\ 
\vspace{0.1cm}
\hspace{-0.2cm} $\sigma_{r}$ [100 ppm] &\hspace{-0.3 cm}  $248 \pm 8$ & \hspace{-0.45cm} --\\ 
\vspace{0.1cm}
\hspace{-0.2cm} $\sigma_{w}$ [100 ppm] &\hspace{-0.3 cm}  $4.956 \pm 0.025$ & \hspace{-0.45cm} --\\ 
\hline
\end{tabular}
\end{table}

\subsection{TOI-674 b}

\begin{figure*}
    \centering
    \includegraphics[width = \textwidth]{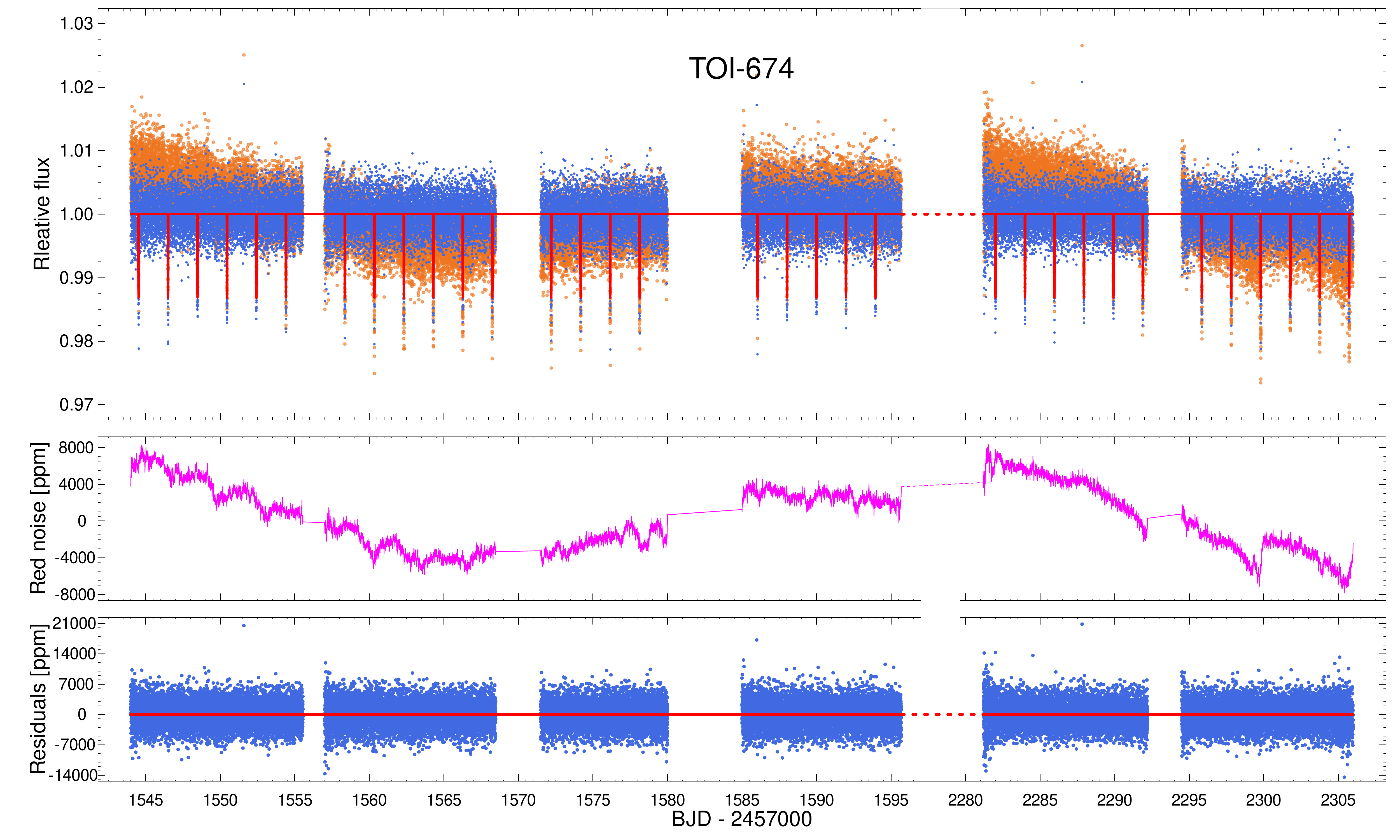}
    \caption{Observed TESS light curve of TOI-674 from Sectors 9, 10, and 36 (top panel, orange dots) -- note the break on the time axis. Subtracting the fitted red noise (middle panel) from the observations yields the blue dots of the top panel. The best fit transit model is shown with continuous red line. The residuals are plotted on the bottom panel.}
    \label{fig:toi674full}
\end{figure*}

\begin{figure}
    \centering
    \includegraphics[width = .48\textwidth]{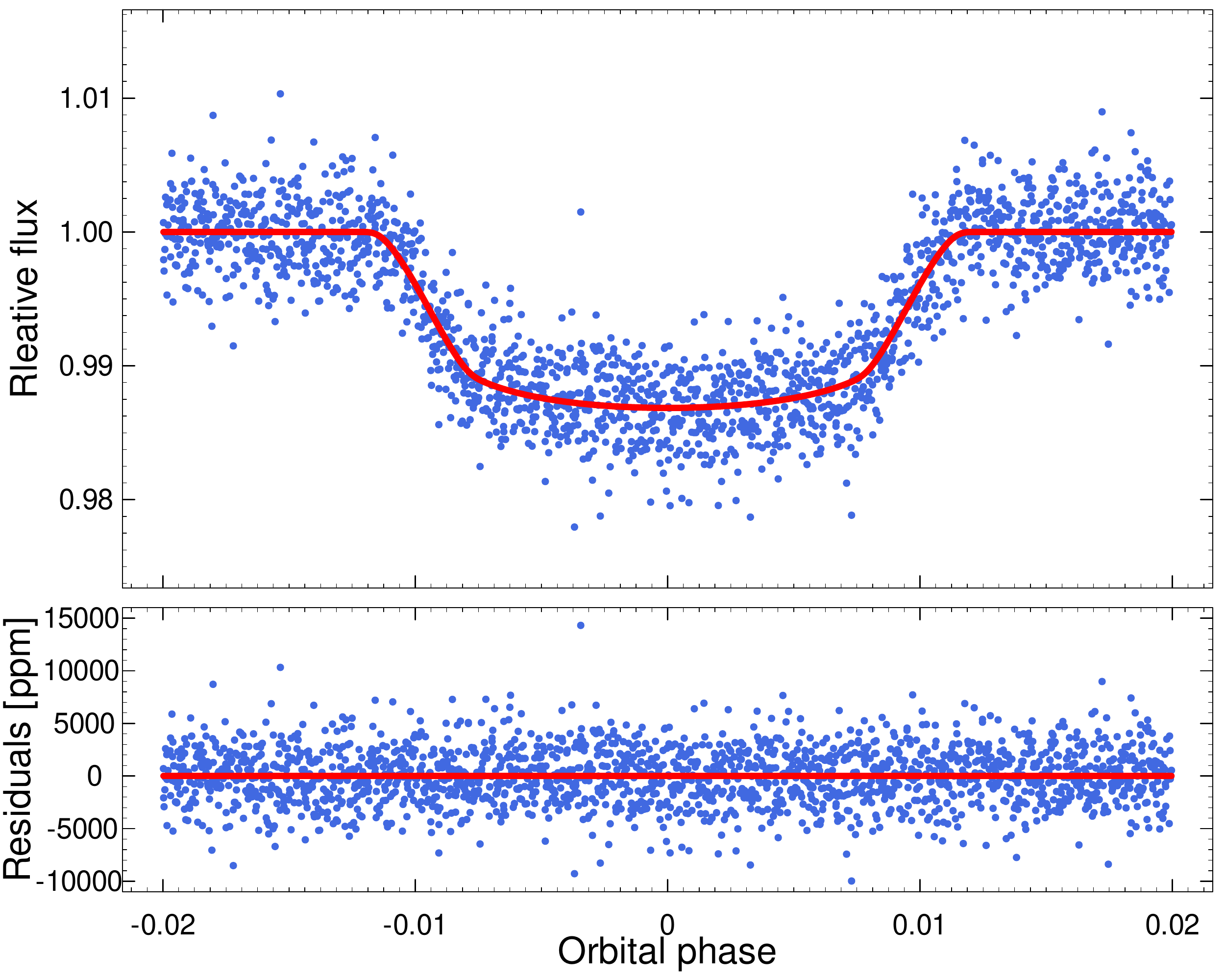}
    \caption{Phase-folded TESS light curve of TOI-674 without correlated noise (top panel, blue dots), the best fit transit model (top panel, solid red line) and the residuals (bottom panel).}
    \label{fig:toi674ph}
\end{figure}

Following \cite{brande}, we re-analyzed the three sectors of \verb|TESS| data with \texttt{TLCM} of TOI-674 b. We adopted the stellar parameters from \cite{murgas} (Table \ref{tab:stellar}) and computed the following limb-darkening coefficients with PyLDTk: $u_{+,\text{TESS}} = 0.59$ and $u_{-, \text{TESS}} = 0.07$. The complete filtered SAP flux light curve is shown on the upper panel of Fig. \ref{fig:toi674full}. The identified red noise is plotted on the middle panel of Fig. \ref{fig:toi674full}; the noise corrected observed light curve; the best fit transit model and the residuals are shown on the upper and lower panels of Figs. \ref{fig:toi674full} and \ref{fig:toi674ph} (as a function of time and orbital phase, respectively). 

Our best-fit transit parameters for TOI-674 b (Table \ref{tab:toi674}) are  $\sim 1\sigma$ from the results of \cite{murgas} for $a/R_S$, $b$, $t_C$ and $P$. In the case of the star-to-planet radius ratio, our results are $2.62\sigma \times$ lower than the previously published values, this is statistically still negligible. The fact that \cite{murgas} find slightly deeper transits for TOI-674 b may be attributed to the fact that we do not include ground-based data and that we use theoretical limb-darkening coefficients.

Based on the wavelets, the white white noise present in the TESS light curve is $438$ ppm $\sqrt{\rm h}$.

\begin{table}
\caption{Comparison of our derived transit parameters for TOI-674b to the results of \citet{murgas}.}
\label{tab:toi674}
\centering
\footnotesize
\begin{tabular}{l c c }
\hline
\hline
Parameter & This work & \cite{murgas} \\
\hline
\vspace{0.1cm}
\hspace{-0.2cm} $a / R_S$ & \hspace{-0.3cm} $12.15 \pm 0.26$ & \hspace{-0.45cm} $12.80 \pm 0.42$\\ 
\vspace{0.1cm}
\hspace{-0.2cm} $R_P / R_S$ & \hspace{-0.3cm} $0.11163 \pm 0.00095$ & \hspace{-0.45cm} $0.1140 \pm 0.0009$\\ 
\vspace{0.1cm}
\hspace{-0.2cm} $b$ & \hspace{-0.3cm} $0.666 \pm 0.017$ &  \hspace{-0.45cm} $0.624 \pm 0.035$\\ 
\vspace{0.1cm}
\hspace{-0.2cm} $P$ [days] & \hspace{-0.3cm}  $1.97716410 \pm 8.4 \cdot 10^{-7}$ & \hspace{-0.45cm} $1.977143 \pm 30\cdot10^{-7}$\\ 
\vspace{0.1cm}
\hspace{-0.2cm} $t_C$ [TBJD] & \hspace{-0.3cm} $1544.52421 \pm 0.00017$ &  \hspace{-0.45cm} $1641.40455 \pm 0.00010$\\ 
\vspace{0.1cm}
\hspace{-0.2cm} $\sigma_{r}$ [100 ppm] & \hspace{-0.3cm} $1003 \pm 25$ & \hspace{-0.45cm} --\\ 
\vspace{0.1cm}
\hspace{-0.2cm} $\sigma_{w}$ [100 ppm] & \hspace{-0.3cm} $24.083 \pm 0.101$ & \hspace{-0.45cm} --\\ 
\hline
\end{tabular}
\end{table}

\subsection{WASP-156 b}

\begin{figure*}
    \centering
    \includegraphics[width = \textwidth]{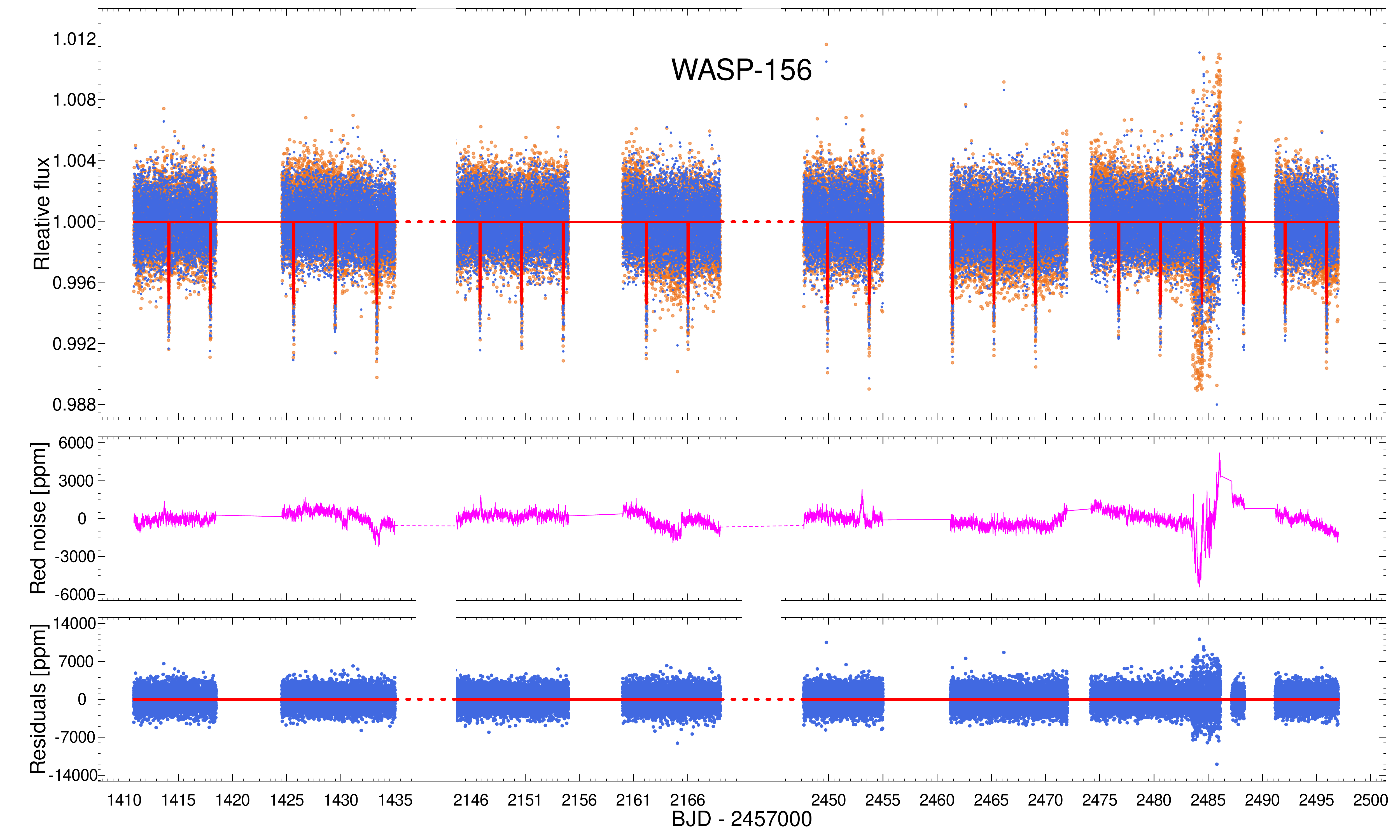}
    \caption{Observed TESS light curve of WASP-156 from Sectors 4, 31, 42 and 43 (top panel, orange dots) -- note the breaks on the time axis. Subtracting the fitted red noise (middle panel) from the observations yields the blue dots of the top panel. The best fit transit model is shown with continuous red line. The residuals are plotted on the bottom panel.}
    \label{fig:wasp156lc}
\end{figure*}

\begin{figure}
    \centering
    \includegraphics[width = .48\textwidth]{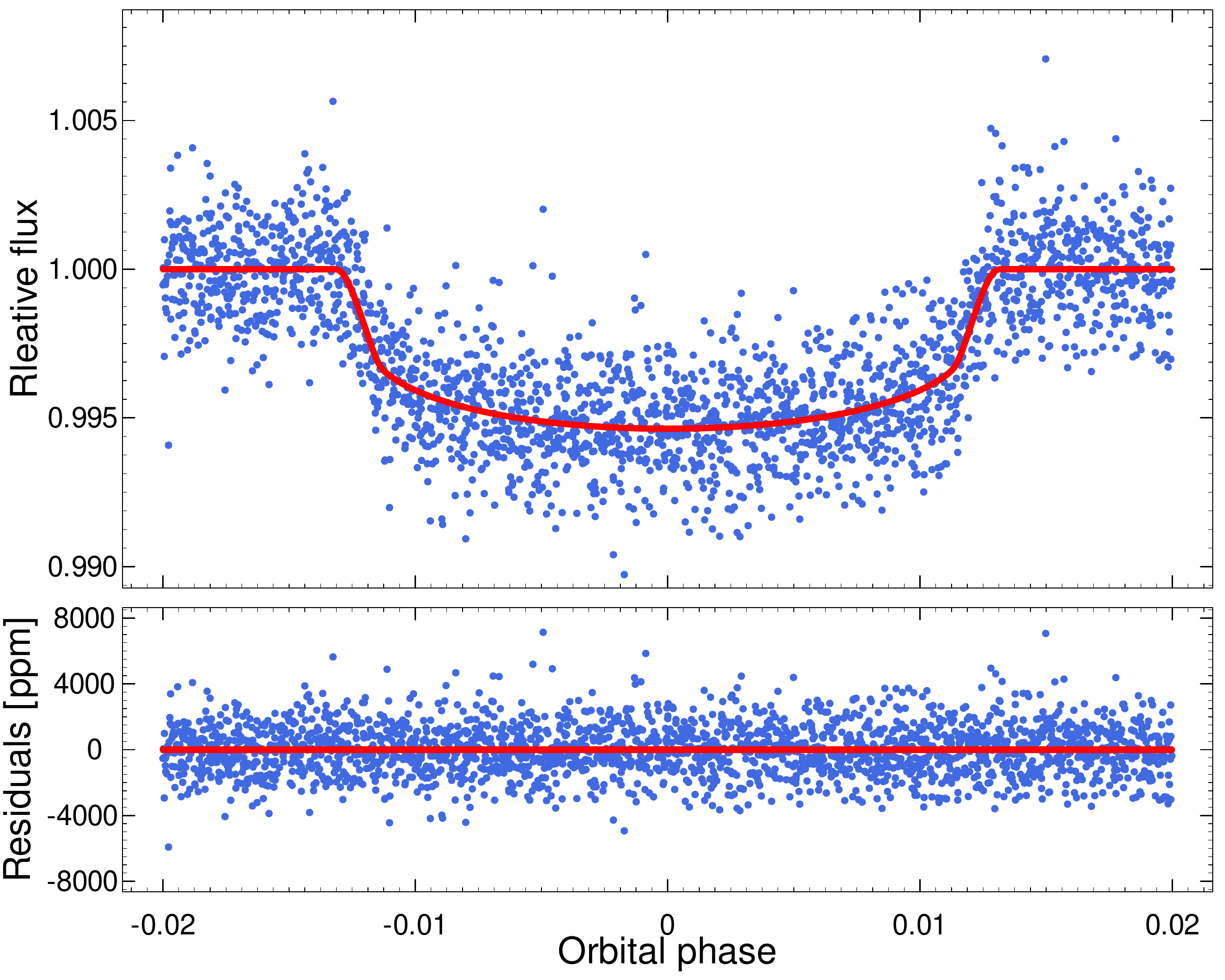}
    \caption{Phase-folded TESS light curve of WASP-156 without correlated noise (top panel, blue dots), the best fit transit model (top panel, solid red line) and the residuals (bottom panel).}
    \label{fig:wasp156ph}
\end{figure}

Our analysis of WASP-156 b was based on the combined SAP flux light curves from S4, S31, S42 and S43 (Fig. \ref{fig:wasp156lc}, top panel). We adopted the stellar parameters from \cite{demangeon} (Table \ref{tab:stellar}) using which the following limb-darkening coefficients were computed by PyLDTk: $u_{+, \text{TESS}} = 0.62$, $u_{-,\text{TESS}} = 0.42$. The best-fit transit model is shown on the upper panels of Figs. \ref{fig:wasp156lc} and \ref{fig:wasp156ph}; the residuals (after the removal of the red noise, shown on the middle panel of Fig. \ref{fig:wasp156lc}) are plotted on the lower panels of Figs. \ref{fig:wasp156lc} and \ref{fig:wasp156ph}.

The fitted transit parameters are shown in Table \ref{tab:wasp156}. All of these are in good agreement with the previous results of \cite{demangeon, 2021AJ....162..221S} \cite{yangfan}. Although our determined uncertainties are considerably larger than the ones given by \cite{2021AJ....162..221S}, \cite{2022arXiv220801716K} showed that the TLCM approach of error estimation yields consistent results.

\begin{table}
\caption{Comparison of our derived transit parameters for WASP-156 to the results of \citet{2021AJ....162..221S}.}
\label{tab:wasp156}
\centering
\footnotesize
\begin{tabular}{l c c }
\hline
\hline
Parameter & This work &  \cite{2021AJ....162..221S}\\
\hline
\vspace{0.1cm}
\hspace{-0.2cm} $a / R_S$ & \hspace{-0.3cm} $12.83 \pm 0.33$ & \hspace{-0.45cm}$12.748^{+0.025}_{-0.027}$\\ 
\vspace{0.1cm}
\hspace{-0.2cm} $R_P / R_S$ & \hspace{-0.3cm} $0.06625 \pm 0.00087$ & \hspace{-0.45cm}$0.067654^{+0.000082}_{-0.000060}$\\ 
\vspace{0.1cm}
\hspace{-0.2cm} $b$ & \hspace{-0.3cm} $0.18 \pm 0.13$ &  \hspace{-0.45cm}$0.2445^{+0.0061}_{-0.0073}$ \\ 
\vspace{0.1cm}
\hspace{-0.2cm} $P$ [days] & \hspace{-0.3cm}  $3.8361604 \pm 2.1\cdot10^{-6}$ & \hspace{-0.45cm}$3.8361603 \pm 4.8 \cdot 10^{-7}$\\ 
\vspace{0.1cm}
\hspace{-0.2cm} $t_C$ [TBJD] & \hspace{-0.3cm} $1414.13598 \pm 0.00046$ &  \hspace{-0.45cm}$1414.136153 \pm 0.000065$ \\ 
\vspace{0.1cm}
\hspace{-0.2cm} $\sigma_{r}$ [100 ppm] & \hspace{-0.3cm} $640 \pm 17$ & \hspace{-0.45cm}\\ 
\vspace{0.1cm}
\hspace{-0.2cm} $\sigma_{w}$ [100 ppm] & \hspace{-0.3cm} $14.490 \pm 0.078$ & \hspace{-0.45cm} \\ 
\hline
\end{tabular}
\end{table}

\section{Prospects with Ariel and PLATO: observing the giraffes} \label{sec:arielplato}

An even better understanding of how the sub-Jovian savanna is formed can be achieved, from an observational point of view, in two ways: precise determination of planetary parameters for many exoplanets (by e.g. the PLATO mission) or bespoke observations for several candidates (by e.g. the James Webb Space Telescope or \verb|Ariel|). We performed simulated observations with PLATO and \verb|Ariel| for the three giraffes described in Sec. \ref{sec:meth} to see how we can improve our understanding of the savanna by refining the transit parameters with these two upcoming ESA missions.

\input{arielparams_table.tex}

\input{table_plato.tex}

\begin{figure*}
   \centering
 \includegraphics[width = \textwidth]{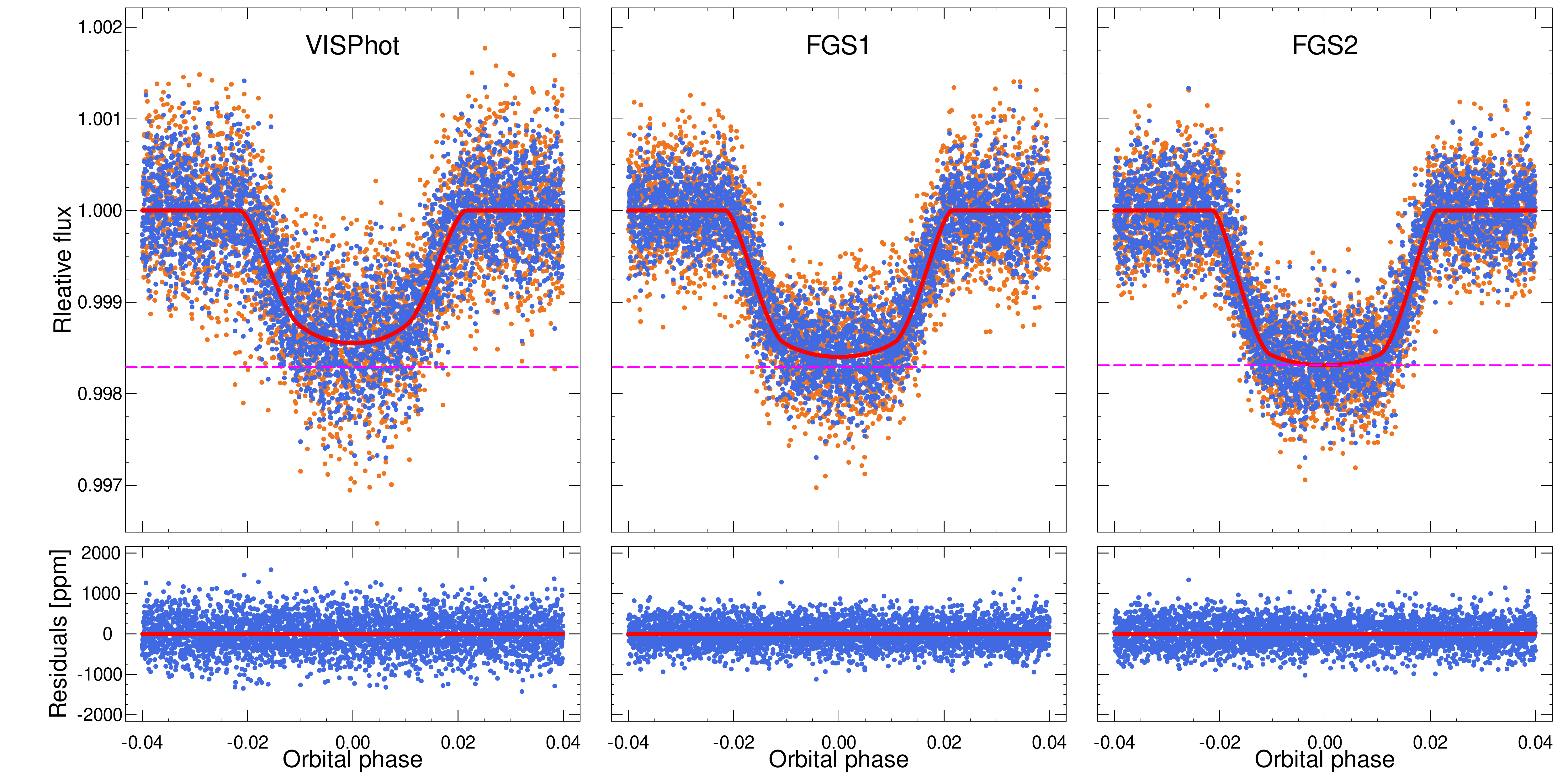}

   \caption{Simulated LTT 9779 b light curves in the three narrow-band filters of Ariel (top panel, orange). Removing the correlated noise yields the blue dots of the top panel. The best-fit transit models are plotted with solid red lines. The residuals are shown in the bottom row. For easier visual differentiation between the different filters, a dashed magenta line is plotted at the minimum of the FGS2 model curve.}
   \label{fig:ltt9779ariel}
\end{figure*}

\begin{figure*}
   \centering
   \includegraphics[width = \textwidth]{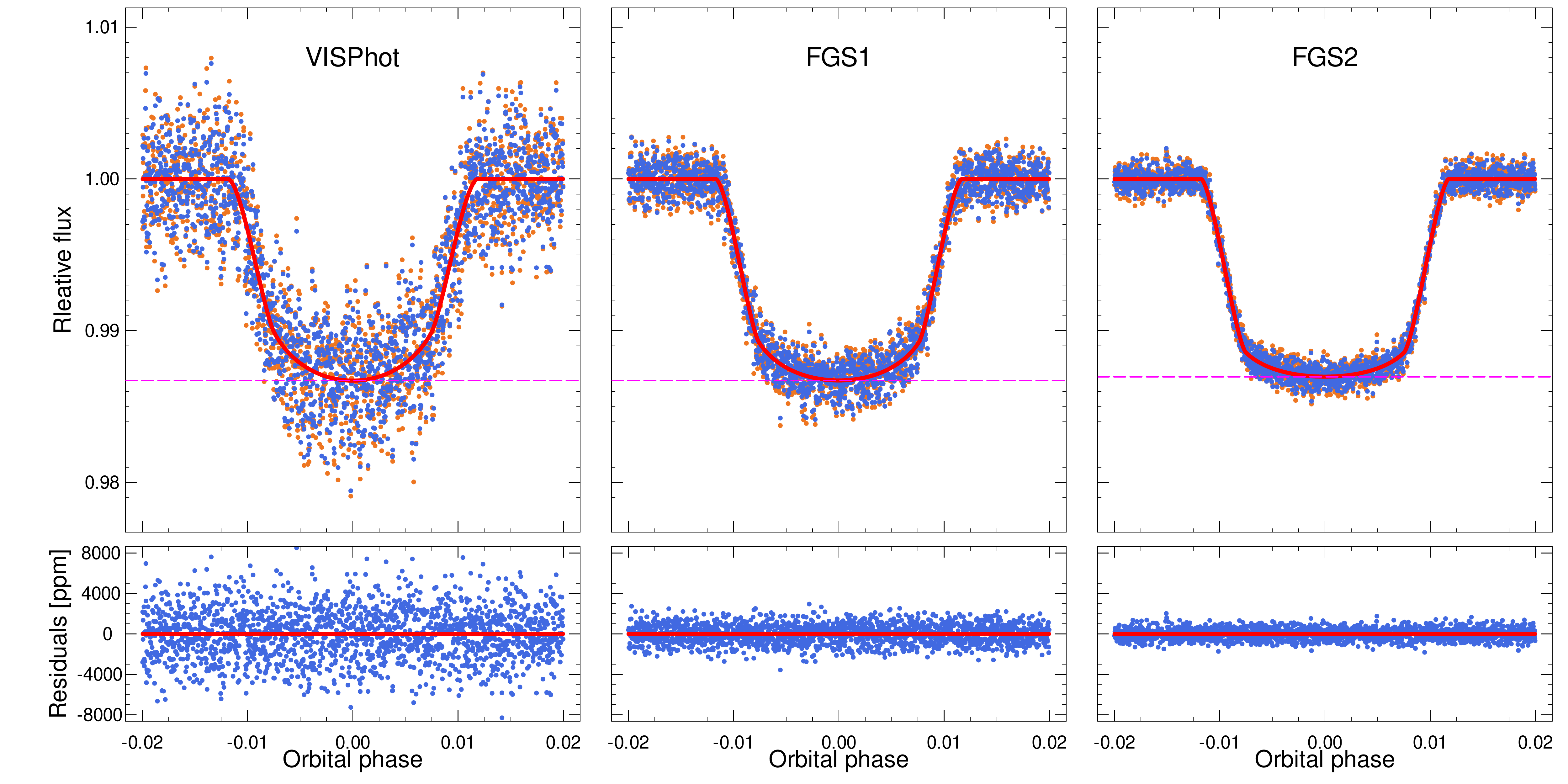}
   \caption{Same as Fig. \ref{fig:ltt9779ariel} but for TOI-674.}
   \label{fig:toi674ariel}
   \end{figure*}

\begin{figure*}
    \centering
    \includegraphics[width = \textwidth]{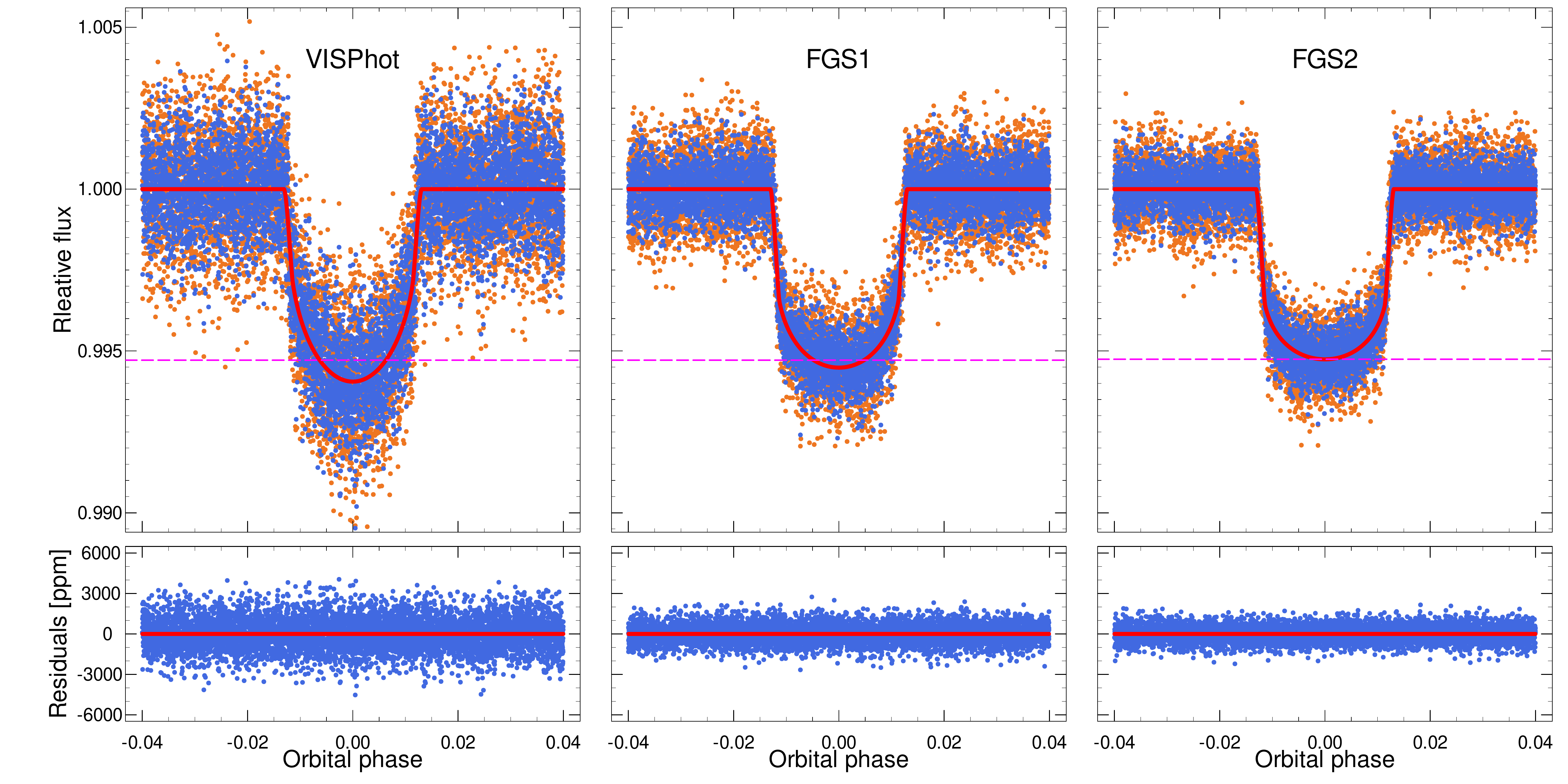}

    \caption{Same as Fig. \ref{fig:ltt9779ariel} but for WASP-156.}
    \label{fig:wasp156ariel}
\end{figure*}

\begin{figure}
    \centering
   \includegraphics[width = 0.48\textwidth]{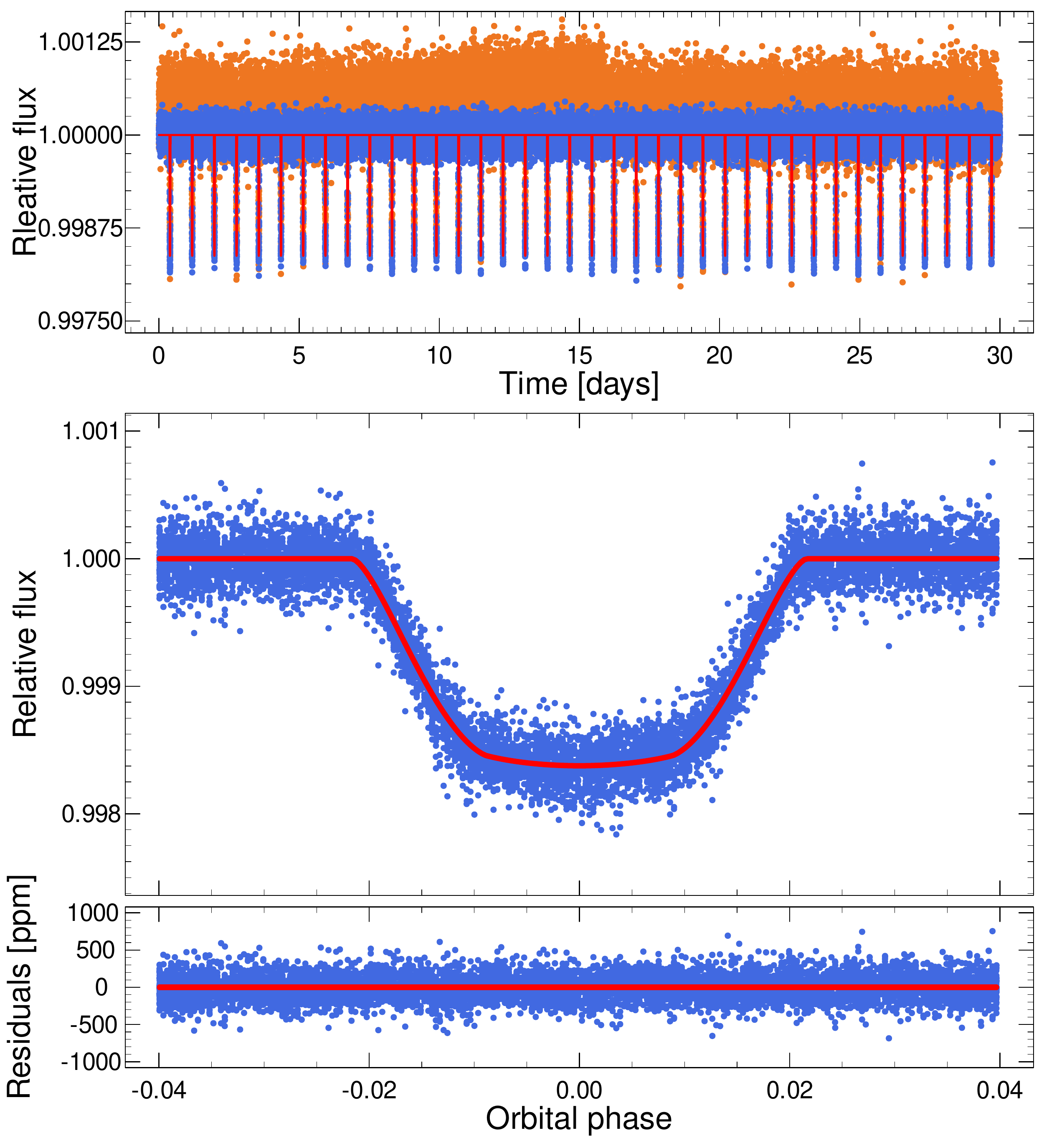}
    \caption{Simulated LTT 9779 observations with $24$ cameras of PLATO (top panel, orange), best-fit transit model (top panel -- as a function of time and middle panel -- as a function of orbital phase) and the light curve with the correlated noise removed (blue dots). The residuals are shown on the bottom panel. For plotting purposes, only every other point is shown.}
    \label{fig:lttplato}
\end{figure}
\begin{figure}
    \centering
    \includegraphics[width = .48\textwidth]{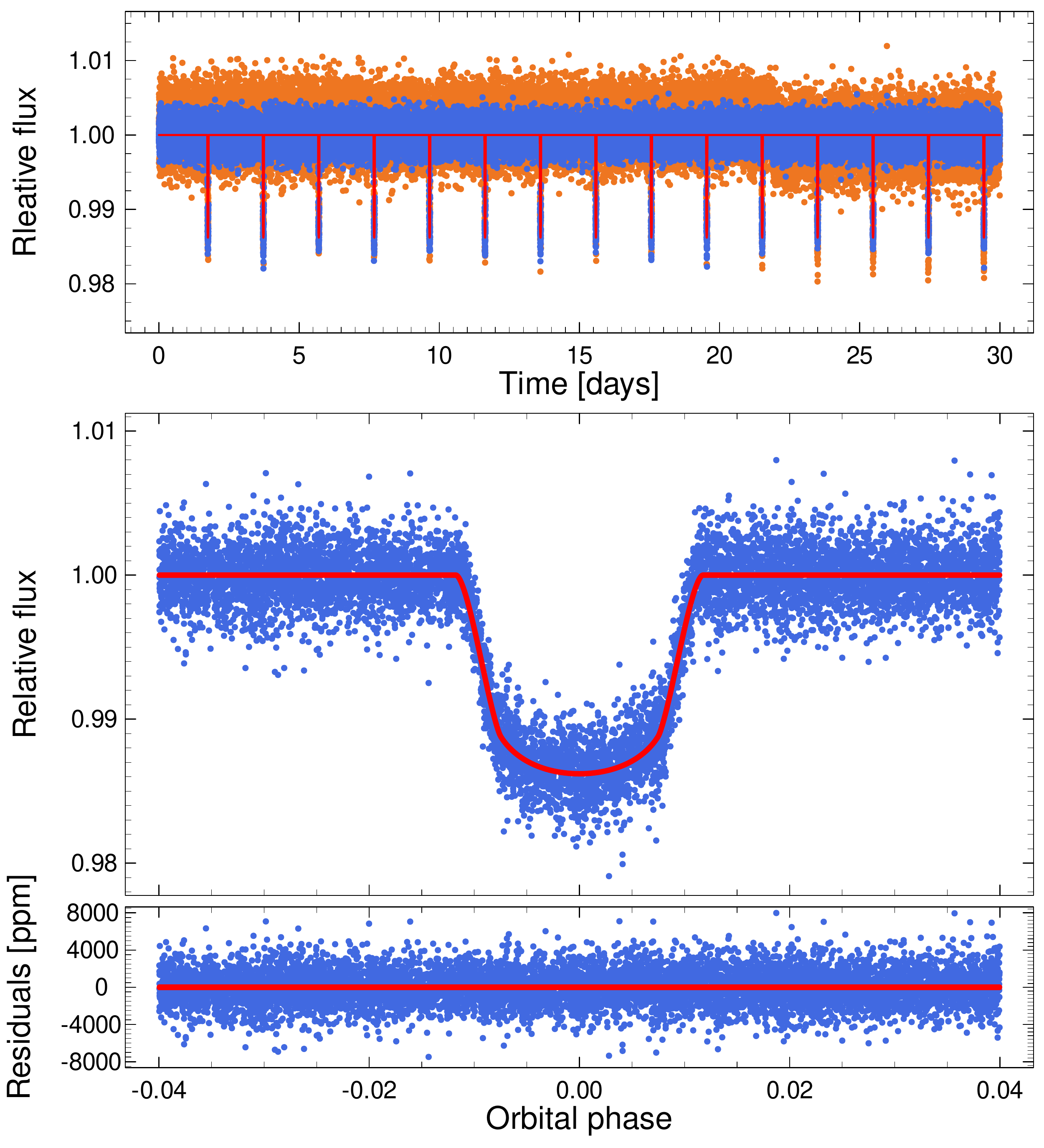}

        \caption{Same as Fig. \ref{fig:lttplato} but for TOI-674.}
    \label{fig:toi674plato}
\end{figure}

\begin{figure}
   \centering
   \includegraphics[width = .48\textwidth]{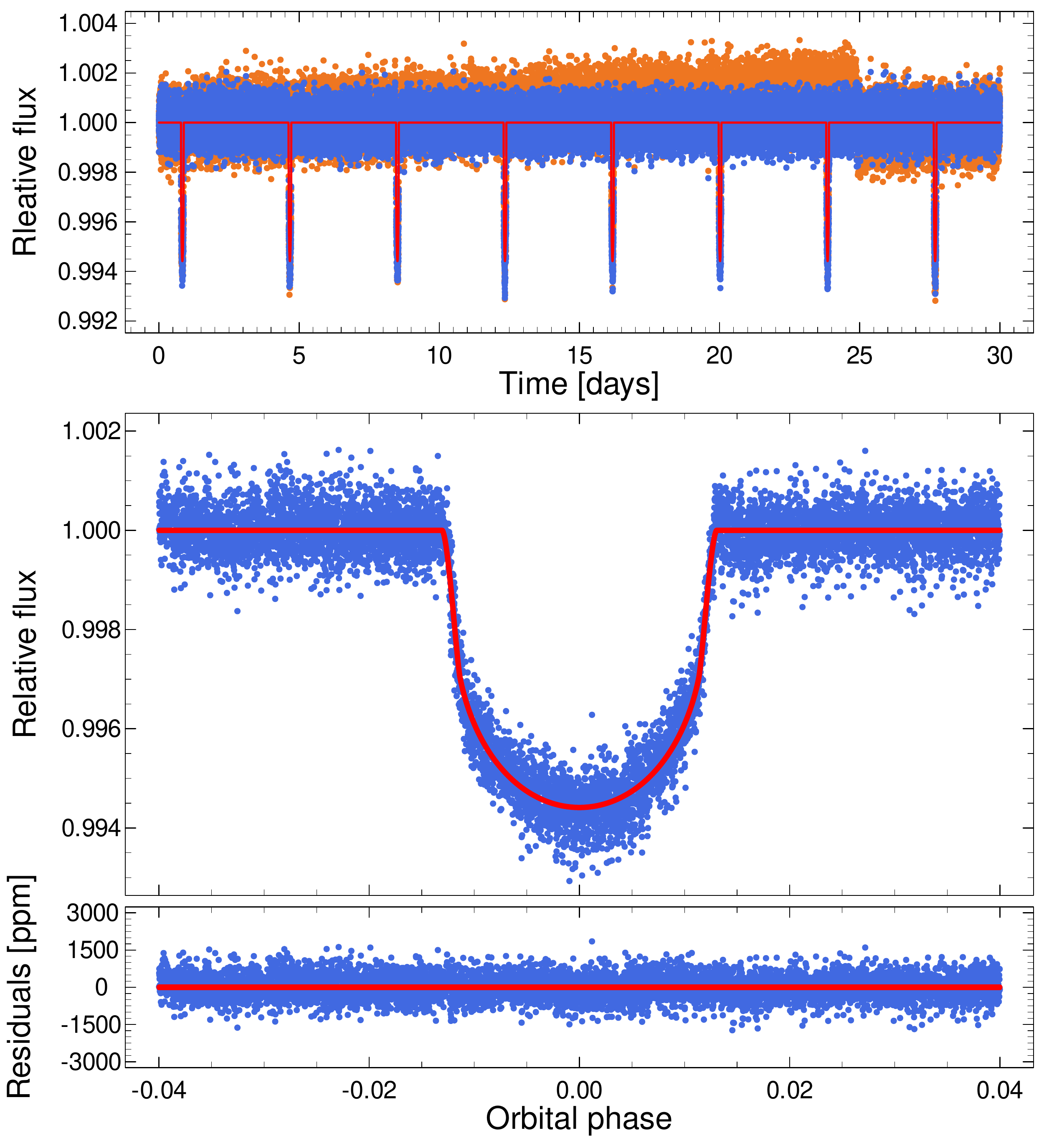}

      \caption{Same as Fig. \ref{fig:lttplato} but for WASP-156.}
   \label{fig:wasp156plato}
\end{figure}

\subsection{Transit parameters from Ariel}

We fitted the simulated \verb|Ariel| light curves in the three filters (VISPhot, FGS1 and FGS2) independently of one another for all three targets.  The phase-folded input light curves and the best-fit models are shown on the upper panels of Figs. \ref{fig:ltt9779ariel}, \ref{fig:toi674ariel} and \ref{fig:wasp156ariel}. The residuals (bottom panels of Figs. \ref{fig:ltt9779ariel}, \ref{fig:toi674ariel} and \ref{fig:wasp156ariel}) indicate that most of the time-correlated noise was removed by the wavelet-formulation incorporated into TLCM. 

The quadratic limb-darkening coefficients were computed with PyLDTk for the three filters and were fixed during the fitting process.The uncertainty in the limb-darkening coefficients was taken into account by fitting $u_+$ and $u_-$ with the inclusion of a Gaussian prior with a central value set according to the theoretical values listed in Table \ref{tab:ldcs} and width of $0.05$ in every case. The other parameters, $P$, $t_C$, $a/RS$, $R_P/R_S$, $\sigma_r$ and $\sigma_w$ were set as free parameters of the fit. As we solve the light curves in the same way for every case, including the \verb|TESS| and PLATO data, we have a homogeneous way of uncertainty estimation, which allows for a better comparison between different filters/instruments. We also note that all uncertainties derived during our analyses correspond to the $1\sigma$ level. Beside the transit parameters from individual filters, the combined values for each parameter was estimated as the weighted mean of the value from the three individual fits, while the combined uncertainty is equal to one-third of the square root of the sum of the individual uncertainties.

\subsubsection{LTT 9779 b with Ariel}
\begin{figure}
    \centering
    \includegraphics[width = 9 cm]{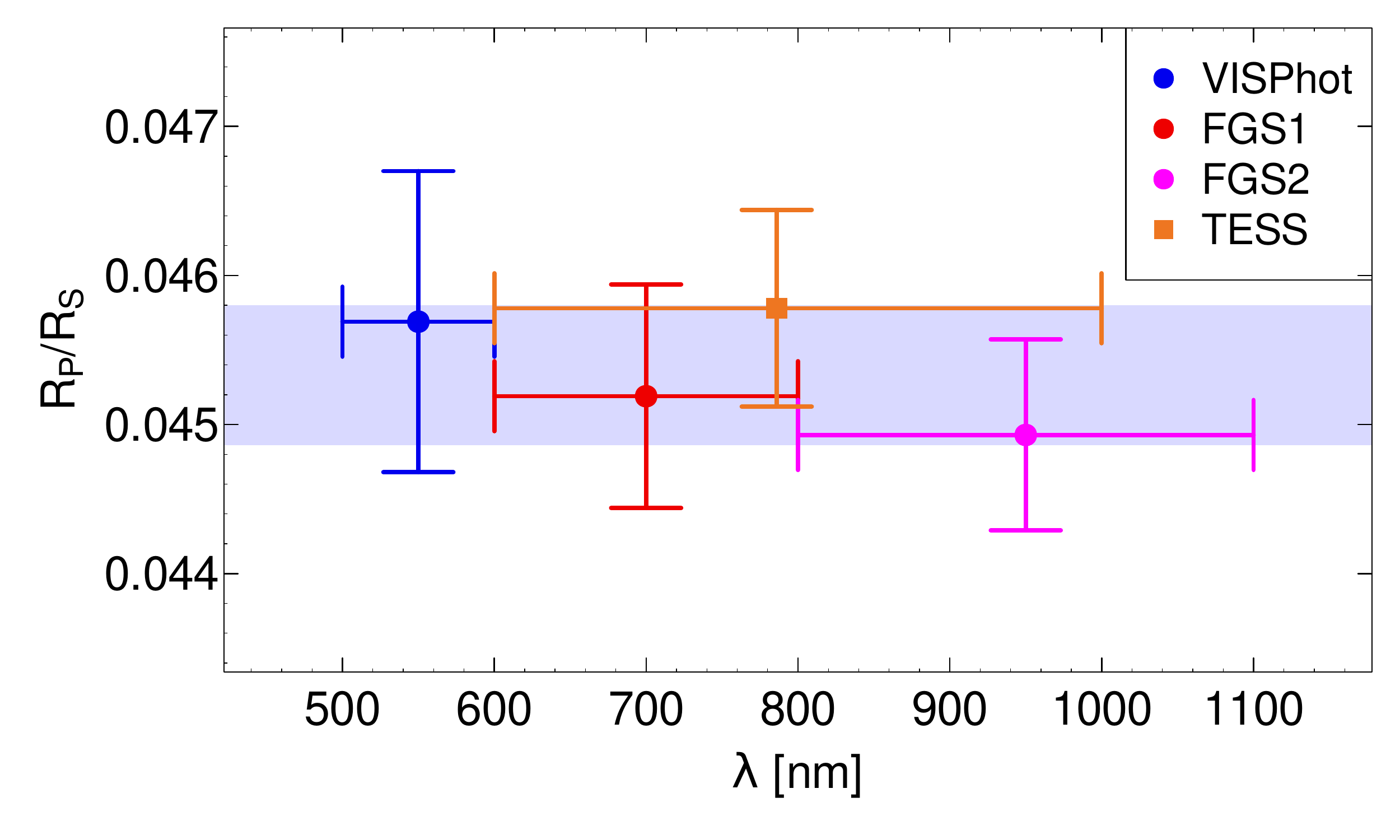}

    \caption{Comparison of the planetary radii fitted in the three bandpasses of Ariel (blue, red, and magenta) and TESS (orange) for LTT 9779 b. The shaded blue area represents the combination of the three photometric filters.}
    \label{fig:ltt9779rads}
\end{figure}
The transit parameters of LTT 9779 from the simulated \verb|Ariel| observations (Fig. \ref{fig:ltt9779ariel}) are shown in Table \ref{tab:giraffes-ariel}. The $t_C$ values are all consistent with $0$, meaning that the correlated noise was removed thoroughly enough for it not to be able to mimic transit timing variations. All other parameters are within $1\sigma$ of the input values, with the exception of $R_P/R_{S, \text{FGS2}}$ at $1.3 \sigma$. Fig. \ref{fig:ltt9779rads} shows the wavelength-(in)dependece of the fitted planetary radii. 
The fitted $\sigma_w$, representing the point-to-point scatter of the white noise in the light curves, are lower in all three filters (at $50$, $37$ and $38$ ppm $\sqrt{\rm h}$) compared to the input values. This apparent discrepancy arises because of two factors: (\textit{i}) the wavelet-based noise filter does not remove all autocorrelated effects from the light curve; it reduces them, and (\textit{ii}) the simple white noise models generated here actually pseudo-random by nature, thus they may include some autocorrelated effects.


The inner precision of the combination of \verb|Ariel| filters during the simulated $30$-day observation is higher 
than what can be computed from the two sectors of \verb|TESS| measurements, with the exception of $P$, the uncertainty of which should improve with a longer time series. Most notably, the planet-to-star radius ratio is with a $\sim 1\%$ precision, which is crucial for estimates of the bulk composition of this planet.

\subsubsection{TOI-674 b with Ariel}

\begin{figure}
    \centering
    \includegraphics[width = 9 cm]{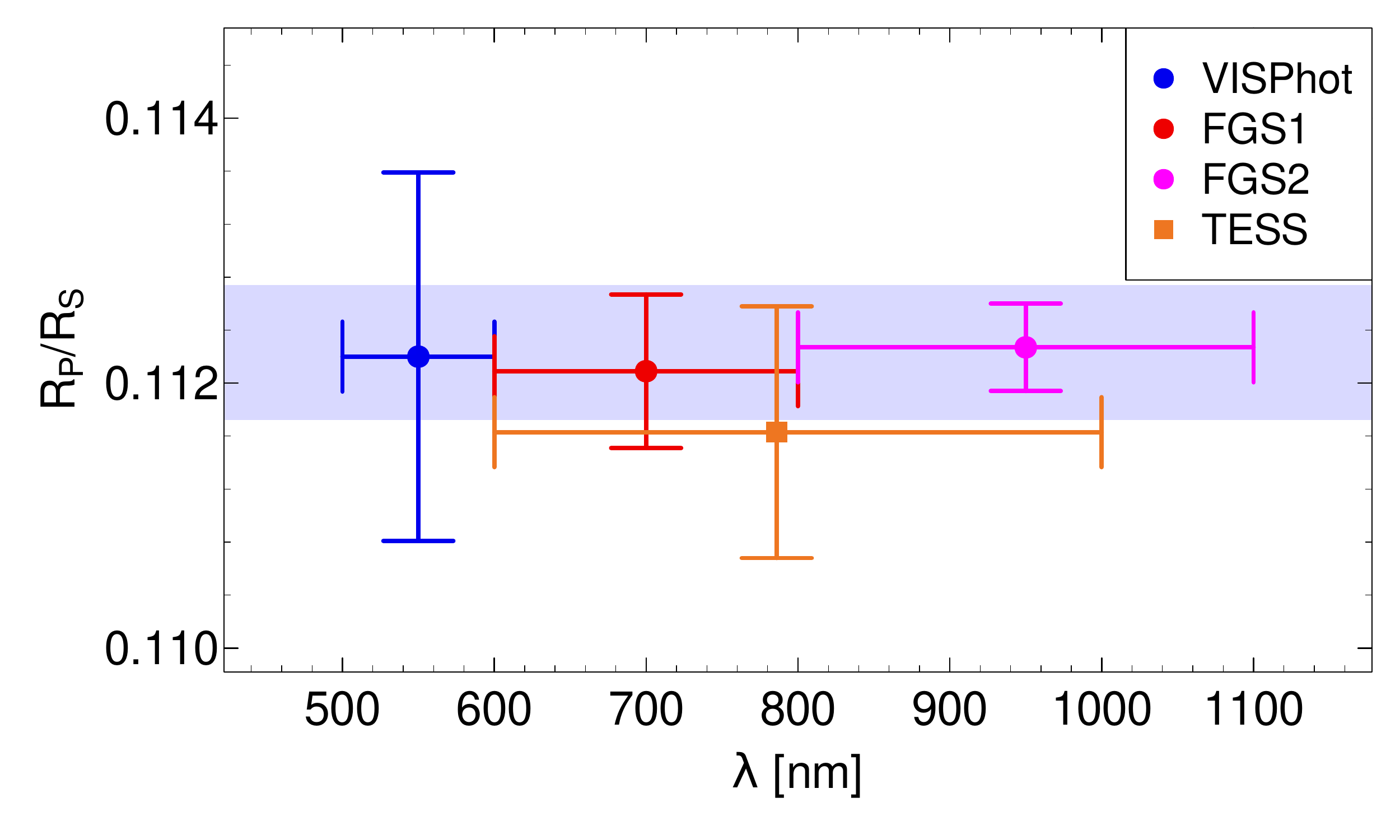}

    \caption{Same as Fig. \ref{fig:ltt9779rads} but for TOI-674 b.}
    \label{fig:toi674rads}
\end{figure}

Simulating one-minute cadence light curves in the three filters of \verb|Ariel| for TOI-674 b (Fig. \ref{fig:toi674ariel}) yields the transit parameters shown in Table \ref{tab:giraffes-ariel}. The fitted transit midtimes are consistent with $0$ in every pass-band. In the case of VISPhot and FGS1, all fitted transit parameters are within $1.3 \sigma$ and $1\sigma$ of their input values, respectively. In FGS2, the retrieved relative planetary radius is higher by $2\sigma$, the semi-major axis is lower by $2.5 \sigma$, and the impact parameter is also higher higher by $2\sigma$. These discrepancies are below the $3\sigma$ threshold for statistical significance, however, they also influence the combined values of the transit parameters, yielding values that are within $2\sigma$ of their respective values derived from the TESS light curves. The fitted planetary radii are plotted as a function of wavelength in Fig. \ref{fig:toi674rads}.



Because of the light curves are solved by TLCM in every case, we may also say that the thirty-day-long simulated \verb|Ariel| observations can improved the precision of the transit parameters in comparison to the three sectors of \verb|TESS| (with the exception of $P$).  Most notably, the $R_P/R_S$ has an relative uncertainty of $0.45\%$ from the combination of the three filters. Furthermore, the fitting process yielded the following estimates for the white noise levels: $274$, $102$ and $59$ ppm $\sqrt{\rm h}$ in VISPhot, FGS1 and FGS2, respectively. As in the case of LTT 9779, these noise levels are lower in comparison with the standard deviation of the input noise models (Table \ref{tab:noise}).

\subsubsection{WASP-156 b with Ariel}

\begin{figure}
    \centering
    \includegraphics[width = 9 cm]{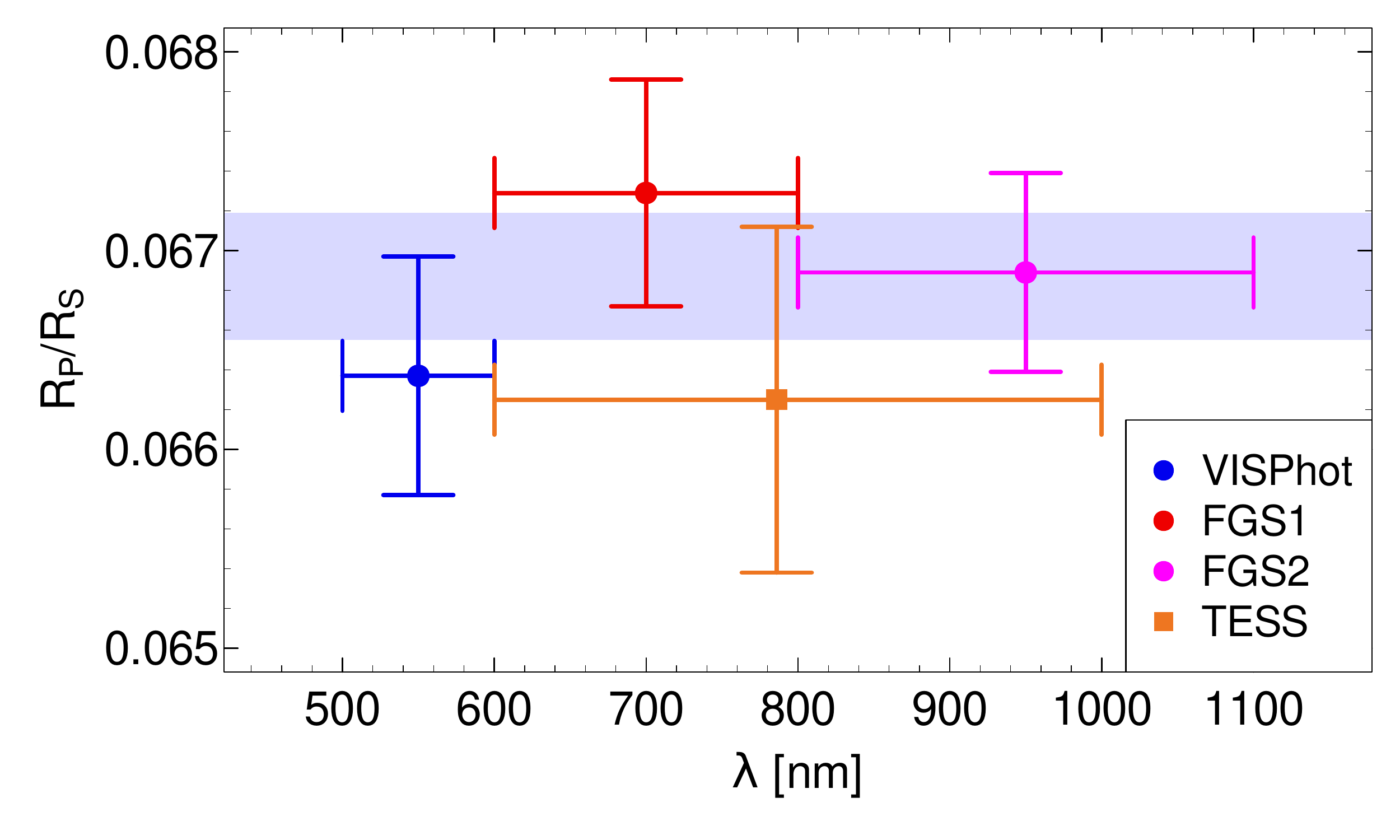}

    \caption{Same as Fig. \ref{fig:ltt9779rads} but for WASP-156 b.}
    \label{fig:wasp156rads}
\end{figure}

The resulting transit parameters from the three narrow-band filters of \verb|Ariel| (Fig. \ref{fig:wasp156ariel}) are shown in Table \ref{tab:giraffes-ariel}. Similarly to the case of LTT 9779 b, the $t_C$ values are consistent with zero in VISPhot and FGS2. All of the fitted transit parameters are within $1.3 \sigma$ from their respective input values in VISPhot and FGS2. In FGS1, the time of midtransit shows a non-zero values at $3 \sigma$, and the orbital period has a $2.8 \sigma$ offset from the derived TESS results, biasing in turn the combined results as well. Because Ariel will observe simultaneously in all three bandpasses, such discrepancies can be handled by excluding the data from the filter where the significant offset in $t_C$ is detected, as it can not have astrophysical origins. The fitted $R_P/R_S$ values for TESS, VISPhot, FGS1 and FGS2 are shown on Fig. \ref{fig:wasp156rads}.

\subsection{Transit parameters from PLATO}

We have selected two 30-day sections of the simulated PLATO light curves where the drift is the most significant (one where the observations are accomplished with all $24$ cameras and one where only $6$ cameras are used). This was done in order to have a better comparison to the \verb|Ariel| simulations and to reduce the computational time. The simulated light curves (for the $24$-camera case) are shown on the top panels Figs. \ref{fig:lttplato}, \ref{fig:toi674plato} and \ref{fig:wasp156plato}, the best-fit transit parameters are shown on the top and middle panels of these figures, while the residuals are plotted on the bottom panels. Similarly to the analysis of the Ariel data, the limb darkening coefficients were fitted by applying a Gaussian prior with a mean set to the theoretical values of Table \ref{tab:ldcs} and a width of $0.05$.

\subsubsection{LTT 9779 b with PLATO}

The light curve from the 24-camera-observations of LTT 9779 is plotted in Fig. \ref{fig:lttplato}; the transit parameters from the two considered cases are shown in Table \ref{tab:giraffes-plato}. The two resulting parameter sets are consistent with each other and are within $1\sigma$ from the input data (obtained form the \verb|TESS| light curves) for $a/R_S$, $R_P/R_S$ and $b$. When we consider the 6-camera-case, the higher noise levels (Table \ref{tab:noise}) induce a $3\sigma$ discrepancy between the input and fitted parameters, while $t_C$ is also not compatible with $0$ ($t_C$ and $P$ are known to be correlated).

Based on $\sigma_w$, the white noise levels of $18$ and $35$ ppm $\sqrt{\rm h}$ are slightly lower than shown in Table \ref{tab:noise}. The planet-to-star radius ratio can be retrieved with a relative precision of $1.1\%$ and $1.6\%$.

\subsubsection{TOI-674 b with PLATO}

Simulated PLATO observations of TOI-674 using all four camera groups are plotted in Fig. \ref{fig:toi674plato}. The retrieved transits are deeper than the input, represented by the $R_P/R_S$ values that are higher by $2.6 \sigma$ and $2.3 \sigma$ in comparison to the TESS data, for the 24 and 6 camera cases, respectively. All other retrieved parameters agree within the estimated uncertainties with their input values. The fitted time of midtransit is also consistent with $0$ in both cases. The nominal precision for the star-to-planet radius ratio is $0.9\%$ and $1.2\%$ for the two considered cases of observations.The fitted $\sigma_w$ values, representing the detected white noise in the system, corresponding to $183$ and $374$ ppm $\sqrt{\rm h}$ for the two considered cases. These values are slightly lower than the level of noise that is superimposed onto the simulated light curves (Table \ref{tab:noise}).

\subsubsection{WASP-156 b with PLATO}

The simulated observations of WASP-156 with PLATO using the more favorable case where the target is observed by all $24$ cameras as well as the best-fit models are plotted in Fig. \ref{fig:wasp156plato}. In both considered scenarios, the fitted transit parameters (Table \ref{tab:giraffes-plato}) are within $1.6\sigma$ of the input values (derived from analyzing the \verb|TESS| light curve). They are also consistent with each other as well as the \verb|Ariel| simulations. The estimated time of midtransit is consistent with $0$. Similarly to the simulated Ariel light curves, the white noise parameter is slightly lower than the standard deviation of the noise on the light curves at $46$ and $91$ ppm $\sqrt{\rm h}$.

One key benefit of the simulations with both upcoming telescopes is that they will present considerable improvements in the precision of the known transit parameters, especially in comparison to the current \verb|TESS| measurements. In the case of WASP-156 b, we can retrieve $R_P/R_S$ with a nominal precision of $1.1\%$ and $1.4\%$ from the $30$-day simulated PLATO observations with the two considered scenarios.

\section{Discussion} \label{sec:disco}
The significance of the presented simulations relies on the ability of the giraffe planets being test-beds for those scenarios of planet formation and evolution which are related to atmospheric mass-loss. The observational evidence point to a multiple structure behind the sub-Jovian desert/savanna \citep{2023arXiv230101065S}. Since its boundaries were proven to be marked by different planets, very probably reflecting different processes, one being more effective in the period--mass parameter spaces, and the other one in the period--radius parameter space. In \cite{2023arXiv230101065S} I, we also discussed that the majority of observations support the irradiation-driven loss of the atmosphere \citep{2018MNRAS.480.2206O, 2019MNRAS.485L.116S}, and the other scenario can likely be the high-eccentricity migration \citep{2018MNRAS.479.5012O}.

We suggest that the significance of the irradiation-driven loss of atmosphere can be directly studied in the case of those planets which are presumably on their way outwards the savanna, and we can reliably assume that they are currently suffering the processes that make the savanna mostly empty. The giraffe planets, being in a location of the parameter space which is otherwise empty, are this kind of test targets. More precisely, there are two scenarios that can retain the giraffe planets inside the savanna: they can either be (1) planets that have recently got into their current position, and due to the processes that empty the savanna, these planets are evolving towards the boundaries within a short time scale, or (2) these are stable positions of these planets, which can be reside inside the savanna for a long time because some specific property protects them from the processes that depose all other planets.

Concerning possibility (2), although planets (or stripped planetary cores) that are dense enough might survive in the savanna such as TOI-849 b \citep{armstrong}, there have been no single parameter or simple combination of parameters foundas of now that are certain to protect the planets inside the savanna, while the time scale of photoevaporation of close-in planets can be as short as one million years \citep{2007A&A...461.1185L}, and hence the explanation (1) seems to be more likely to us. Also, according to \cite{2021A&A...649A..40D}, the boundary of the savanna is populated by several exoplanets with significant atmospheric escape \citep[][and references therein]{2021A&A...649A..40D}. 
However, in either of these cases, the precise determination of planet parameters can reveal the reasons behind the formation of the savanna itself -- either by observing how planets \textit{leave} the savanna, or by observing how the can \textit{survive}.

The photoevaporation of planets is a process which was observed in the case of a few exoplanets, and it's detection --if it really is reshaping the giraffe planets-- is promising. A feature of photoevaporation is a metallicity-dependent thermal cooling \citep{2018A&A...610A..63D, petigura}, which can be revealed by a highly sensitive thermal analysis of the atmosphere (although the effect is superimposed on a complicated thermal map which itself can have prominent asymmetries \citep[e.g.][]{2007Natur.447..183K, 2010ApJ...721.1861A, 2020A&A...643A..94L}. Such observations could be made by analyzing the phase curves and eclipses of the planets, which is more easily done in infra-red due to the temperatures expected on the planet \citep[see e.g.][]{2010ApJ...721.1861A,2018haex.bookE.147C} in combination with spectroscopic observations which can be used to probe the atmosphere directly. This is therefore clearly a task well-suited for Ariel.

Another, and quite routinely observed feature linked to irradiation-driven loss of atmosphere is the presence of tenuous gas envelopes around the exoplanets. These features can be detected in the transmission spectra taken during the transits in either atomic lines \citep[e.g.][]{2018Sci...362.1388N} or molecules \citep[e.g.][]{2015Sci...350...64M}, the latter of which will be possible with the Ariel InfraRed Spectrometer (see \cite{2020AJ....160...80C} for details). A photometric feature related to the atmospheric envelope is the wavelength-dependent transit depth, caused by the molecular bands in the IR spectrum, and the wavelength-dependence of the total scattering cross section of a planet with a chemical gradient in an extended upper atmosphere/ionosphere \citep[e.g.][]{2020AJ....160....8E}, with a spectacular example being the detection of CO$_2$ in the atmosphere of WASP-39 b using \verb|JWST| \citep{2022arXiv220811692T}. In the case of TOI-674 b, \cite{murgas} attempted to constrain the wavelength-dependent transit depth based on \verb|Spitzer| and \verb|HST| measurements and found evidence of water vapor, a feature that could be observed by \verb|Ariel| with higher significance.

A third observation can be the time-dependece of $R_p$, due to the loss of material and the destruction of an overinflated atmosphere, which can appear with a transit light curve atmosphere in the case of material loss with a comet-like intensity \citep{2012ApJ...752....1R,2018A&A...611A..63G}. Variability in the transit depth and shape may occur on timescales as short as weeks--months as in the case of K2-22 b \citep{2015ApJ...812..112S}. As both wavelength and time-dependence of $R_p$ can be expected, a baseline planetary radius is necessary which can not be reliably determined from observations at a singular passband, or from IR spectrophotometry. Our proposed observing scenario can therefore only be carried out via Ariel, whereas the chief purpose of PLATO observations would be monitoring the time-dependence.

In case of future JWST observations (a facility that would of course be capable of completing a similar, though not identical, program) of these specific targets, follow-up Ariel observations would be even more crucial for a simultaneous search for time and wavelength-dependence. However, contrary to JWST, the primary focus of Ariel will be observations of exoplanetary systems. Given that the key for higher precision would be observations of many transits and not the size of the primary mirror, Ariel is more suitable for such a program.

Observations with the IR spectrograph of Ariel would also add a great deal of new information to the presented ideas. To complete such simulations, introduction of many unknown atmospheric parameters would be mandatory. The best exploitation of spectrophotometry with Ariel will be explored in Szab\'o et al (in prep).

These observations have a very promising future prospects in the forthcoming years. The discovery of new giraffe planets can be expected from \verb|PLATO| \citep{2014ExA....38..249R}, reaching a fainter detection limit than TESS \citep{ricker}. Due to the short orbital period of these planets, an observation time covering approximately one \verb|TESS| Sector ($\approx$ 28 days with gaps) will be enough for a secure detection (and this is why we expect \verb|TESS| already have  discovered most of the giraffe planets within its observable sky). The observational history of transit depths of exoplanets in optical bands, including but not limited to the \verb|Kepler| \citep{2010Sci...327..977B}, \verb|TESS|, \verb|CHEOPS| \citep{2021ExA....51..109B}, and being expected from \verb|PLATO| and \verb|Ariel| \citep{2021arXiv210404824T} will cover 2 decades for many exoplanets, leading to precise radii. Transmission spectra with the James Webb Space Telescope will be also a standard tool for atmospheric studies in exoplanet \citep[e.g.][]{2022ApJ...928L...7R}. Also, time evolution of $R_p/R_s$ or asymmetries in the transit light curve can reveal atmospheric evaporation or disintegration (although currently known transit depth variations are linked to nodal precession instead \citep[e.g.][]{2012MNRAS.421L.122S, 2020MNRAS.492L..17S}, chromospheric activity \citep{2008ApJ...682..593M, 2010ApJ...721.1861A} or inhomogeneities of the stellar surface \citep{2022A&A...659L...7S}). 

There are also several other systems which are argued to be in the sub-Jovian savanna (including NGTS-4 b, NGTS-10 b, TOI-2196 etc.), and as we argued in \cite{2023arXiv230101065S}, the savanna itself has no firm border and has a multitude of planets near it. It is also reasonable to expect that there are other unknown or uncharacterized (a task suited for PLATO) such planetary systems. Therefore even if these specific targets are not observed by Ariel, direct observational evidence for the appearance of the savanna can be gained from measurements of many other candidates similar to the giraffes explored in this paper. This paper should be regarded as a proof of concept instead of a direct definition of a specific study.

\section{Summary and conclusion}

The area in the $R_P$ -- $P$ and $M_P$ -- $P$ parameter spaces formerly known as the sub-Jovian/Neptune desert \citep{2019MNRAS.485L.116S} is by now known to be not entirely empty \citep{2020NatAs...4.1148J, murgas} hence we proposed  its conversion into the ``sub-Jovian savanna''. In this paper, we selected three known exoplanets that are either in the sub-Jovian savanna (such as LTT 9779 b and TOI-674 b), or are on its border (such as WASP-156 b) to explore how future the two future ESA missions (\verb|PLATO| and \verb|Ariel|) might help resolve the complex processes behind the formation of the savanna (see \cite{2018MNRAS.479.5012O} and \cite{2023arXiv230101065S} for details).

As the selected objects inhibit the ``savanna'', we also coined the term giraffe planets. First we analysed all available \verb|TESS| data of these giraffes (two, three, and four Sectors for LTT 9779, TOI-674 and WASP-156, respectively) with \texttt{TLCM}. In general, our transit parameters are in good agreement with the previously published results of \cite{2020NatAs...4.1148J, 2021AJ....162..221S, murgas}. Our analysis of the \verb|TESS| light curves improved the precision of these parameters, which is the result of the inclusion of more data and the wavelet-based noise handling of \texttt{TLCM}. 
 
We used the transit parameters derived from the \verb|TESS| analysis as input to the simulated \verb|Ariel| and \verb|PLATO| observations. In the case of \verb|Ariel|, we devised an algorithm to simulated the most realistic observations using the three narrow passbands (VISPhot, FGS1 and FGS2). The transits (simulated with the \texttt{batman} software package) were injected into a time series consisting of white noise (based on \texttt{ArielRad} calculations) and red noise (based on Kepler light curves of stars similar to the hosts) which included both stellar and instrumental noise sources. In the case of the simulated PLATO observations, we made use of \texttt{PSLS} to create PLATO-like noise that does not include any astrophysical effects and injected the transits into it. As the PLATO fields are not yet fixed, we have no knowledge of how these targets might be observed or if they will be observed at all. We therefore considered the most and least favorable cases (observations with all $24$ cameras and with just six camera). It is also reasonable to expect that in case these exact targets are not observed at all, other giraffe planets should be discovered before the launch of PLATO and during its operation. 

The simulated light curves were solved with \texttt{TLCM}. In the case of the three \verb|Ariel| filters, the light curves were solved independently and the parameters were then combined, yielding constrains on the star-to-planet ratio with precision $\sim 1\%$. In the case of \verb|PLATO|, the selected $30$-day-long light curves resulted in transit parameters estimated with higher precision than the currently available \verb|TESS| data, in both considered cases. The estimated uncertainties for the \verb|PLATO| simulations were slightly higher than the results of the combined \verb|Ariel| observations. We elected not to include astrophysical noise sources in the \verb|PLATO| simulations for the following reasons. (\textit{i}) by adding more noise, the uncertainties of the transit would be at least as large as the ones estimated in Table \ref{tab:giraffes-plato}, which are already less precise than the \verb|Ariel| data. (\textit{ii}) a direct comparison between the light curves obtained from the three different telescopes will not be possible for the any real observations due to the different approaches for observations of each facility. If \verb|PLATO| were to observe these specific targets, we would have light curves spanning at least $90$ days \citep{nascimbeni}, which would ultimately yield higher precision than these simulations. (\textit{iii}) allocating time for any targets will be more challenging with Ariel, as it will observe $\approx 1000$ targets \citep{2021arXiv210404824T}, also, the simultaneous photometric and spectroscopic measurements will only be possible by \verb|Ariel|, thus a more thorough analysis of the feasibility of these observations is warranted. (\textit{iv}) if there will be observations of \verb|PLATO| of these targets, those would serve as extraordinary follow-up measurements to the pre-existing \verb|TESS| data, regardless of the actual precision of the transit parameters.
 
The giraffe planets can be golden targets for the \verb|Ariel| mission, because the simultaneous photometry and spectroscopy enables the comparison of the transit depth in the visual and infrared bands, which is a possibility to immediately recognise atmospheric processes related to loss of atmosphere. The measurements can be repeated, and the dynamics or possible variations in evaporation rates will be also revealed. In atmospheric studies, the planet radius is an important input parameter which appears in several mutual parameter degenerations. The most sensitive detection of the atmospheric loss requires an accurate, prior information about the planet radius, while the radius itself is degenerated with stellar limb-darkening coefficients as well \citep{2013A&A...549A...9C}.  As the disintegration of these planets can be expected, changes in the transit shape would introduce further degeneracies between the transit parameters. Ariel will be capable of a simultaneous observation of the transit light curve in 3 optical bands and break up the degeneracy between the planet radius and the limb-darkening. This will lead to an unbiased and precise radius determination \citep{2022ExA....53..607S} and together with the simultaneously taken transmission spectra in the infrared \citep{2021arXiv210404824T}, the structure of the upper atmosphere will be revealed. We suggest that the homogeneous studies of the giraffe planets with Ariel should be a key study in the formation and evolution of close-in planets.

\section*{Data availability}
The simulated Ariel and PLATO light curves are available in Open Science Framework, at \url{https://osf.io/vyrx9/?view_only=a2bf4b774f884ac4866af2c1436393d9}.

\section*{Acknowledgements}

The authors thank M. Bergemann and G. Morello for their valuable comments.  This work was supported by the PRODEX Experiment Agreement No. 4000137122 between the ELTE E\"otv\"os Lor\'and University and the European Space Agency (ESA-D/SCI-LE-2021-0025). Support of the Lend\"ulet LP2018-7/2021 grant of the Hungarian Academy of Science, and the KKP-137523 `SeismoLab' \'Elvonal  grant  as well as the grant K-138962 of the Hungarian Research, Development and Innovation Office (NKFIH) are acknowledged. LBo acknowledges the funding support from Italian Space Agency (ASI) regulated
by “Accordo ASI-INAF n. 2013-016-R.0 del 9 luglio 2013 e integrazione del 9 luglio 2015”.  Project no. C1746651 has been implemented with the support provided by the Ministry of Culture and Innovation of Hungary from the National Research, Development and Innovation Fund, financed under the NVKDP-2021 funding scheme.

This work presents results from the European Space Agency
(ESA) space mission PLATO. The PLATO payload, the PLATO
Ground Segment and PLATO data processing are joint developments of ESA and the PLATO Mission Consortium (PMC).
Funding for the PMC is provided at national levels, in particular
by countries participating in the PLATO Multilateral Agreement
(Austria, Belgium, Czech Republic, Denmark, France, Germany,
Italy, Netherlands, Portugal, Spain, Sweden, Switzerland, Norway,
and United Kingdom) and institutions from Brazil. Members of the
PLATO Consortium can be found at https://platomission.com/. The
ESA PLATO mission website is https://www.cosmos.esa.int/plato.
We thank the teams working for PLATO for all their work

\bibliographystyle{mnras}\bibliography{refs}

\bsp	
\label{lastpage}
\end{document}

%% file: noise_table.tex
\begin{table*}
\caption{\textbf{White and red noise levels used in the simulations for the three targets in the three Ariel pass-bands (VISPhot, FGS1 and FGS1), the two considered cases for the PLATO visits ($6$ and $24$-camera observations) and the noise levels for the Kepler data, expressed as point-to-point scatter in one-hour bins. These values are calculated as the standard deviation of each noise model (see text for more details). Note that the Kepler noise levels are calculated for the stars which were chosen as the source for the red noise.}}
\label{tab:noise}
    \centering
    \scriptsize
    \begin{tabular}{l c c c c c c c c c c c c c c c c c c}
    \hline
    \hline
   Parameter  & \multicolumn{6}{c}{LTT 9779} &  \multicolumn{6}{c}{TOI-674} &  \multicolumn{6}{c}{WASP-156}  \\
    \hline
    & \hspace{-0.2cm}VISphot & \hspace{-0.2cm}FGS1 & \hspace{-0.2cm}FGS2 & \hspace{-0.2cm}PLATO 24 & \hspace{-0.2cm}PLATO 6&\hspace{-0.2cm} Kepler  & \hspace{-0.2cm}VISphot & \hspace{-0.2cm}FGS1 & \hspace{-0.2cm}FGS2 & \hspace{-0.2cm}PLATO 24 & \hspace{-0.2cm}PLATO 6 & \hspace{-0.2cm}Kepler   & \hspace{-0.2cm}VISphot & \hspace{-0.2cm}FGS1 & \hspace{-0.2cm}FGS2 &\hspace{-0.2cm} PLATO 24 & \hspace{-0.2cm}PLATO6 & \hspace{-0.2cm}Kepler \\
    WN [ppm $\sqrt{\rm h}$] & \hspace{-0.2cm}$66$ & \hspace{-0.2cm}$50$& \hspace{-0.2cm}$47$ &\hspace{-0.2cm} $26$ & \hspace{-0.2cm}$49$ &\hspace{-0.2cm}$286$   & \hspace{-0.2cm}$354$ & \hspace{-0.2cm}$132$ &\hspace{-0.2cm} $78$ &\hspace{-0.2cm} $230$ & \hspace{-0.2cm} $434$ &\hspace{-0.2cm} $310$  &\hspace{-0.2cm} $145$ &\hspace{-0.2cm} $83$ &\hspace{-0.2cm} $71$ &\hspace{-0.2cm} $61$ &\hspace{-0.2cm}$126$ &\hspace{-0.2cm} $286$ \\
    RN [ppm $\sqrt{\rm h}$]& \hspace{-0.2cm}$126$ &\hspace{-0.2cm} $121$ &\hspace{-0.2cm} $116$ &  \multicolumn{2}{c}{$85$} &\hspace{-0.2cm}$213$    & \hspace{-0.2cm}$120$ & \hspace{-0.2cm}$112$ &\hspace{-0.2cm} $102$ & \multicolumn{2}{c}{$122$} &\hspace{-0.2cm} $208$ &\hspace{-0.2cm}   $143$ &\hspace{-0.2cm} $136$ &\hspace{-0.2cm} $127$ & \multicolumn{2}{c}{$134$} & \hspace{-0.1cm}$265$\\
    \hline
    \end{tabular}
\end{table*}

%% file: arielparams_table.tex
\begin{table*}
\caption{Derived transit parameters for LTT-9779 b, TOI-674 b and WASP-156 b from the three filters of Ariel and the combined values.}
\label{tab:giraffes-ariel}
\centering
\begin{tabular}{l c c c c}
\hline
\hline
Parameter & VISPhot & FGS1 & FGS2 & Combined \\
\hline \\[\dimexpr-\normalbaselineskip+3pt]
\multicolumn{5}{c}{\textit{LTT 9779 b}} \\ [\dimexpr-\normalbaselineskip+12pt]
\hline
$a / R_S$ &  $3.772 \pm 0.041$ & $3.782 \pm 0.040$ & $3.790 \pm 0.044$ & $3.784 \pm 0.024$ \\ 
 $R_P / R_S$ &  $0.04569 \pm 0.00101$ & $0.04519 \pm 0.00075$ & $0.04493 \pm 0.00064$ & $0.04533 \pm 0.00047$ \\ 
 $b$ &  $0.9223 \pm 0.0040$ &  $0.9215 \pm 0.0032$ & $0.9231 \pm 0.0030$ & $0.9223 \pm 0.0020$ \\ 
 $P$ [days] &  $0.792065 \pm 0.000016$ & $0.792067 \pm 0.000011$ & $0.792060 \pm 0.000009$ & $0.792064 \pm 0.000007$ \\ 
 $t_C$ [days] & $0.000104 \pm 0.00034$ & $-0.000064 \pm 0.000237$ & $0.000027 \pm 0.000185$ & $0.000032 \pm 0.000150$ \\ 
 $u_{+}$ &  $0.81 \pm 0.06$ & $0.64 \pm 0.05$ & $0.51 \pm 0.06$ & -- \\ 
 $u_{-}$ &  $0.66 \pm 0.07$ & $0.43 \pm 0.07$ & $0.29 \pm 0.07$ & -- \\ 
$\sigma_{r}$ [100 ppm] & $294.202 \pm 4.058$ & $243.431 \pm 2.885$ & $159.460 \pm 2.714$ & -- \\ 
$\sigma_{w}$ [100 ppm] & $3.854 \pm 0.017$ & $2.897 \pm 0.014$ & $2.924 \pm 0.013$ & --  \\
\hline
\\[\dimexpr-\normalbaselineskip+3pt]
\multicolumn{5}{c}{\textit{TOI-674 b}} \\ [\dimexpr-\normalbaselineskip+12pt]
\hline
$a / R_S$ &  $11.73 \pm 0.28$ & $12.10 \pm 0.16$ & $11.87 \pm 0.11$ & $11.93 \pm 0.11$ \\ 
 $R_P / R_S$ &  $0.11220 \pm 0.00139$ & $0.11209 \pm 0.00058$ & $0.11227 \pm 0.00033$ & $0.11223 \pm 0.00051$ \\ 
 $b$ &  $0.693 \pm 0.021$ &  $0.672 \pm 0.013$ & $0.686 \pm 0.008$ & $0.683 \pm 0.009$ \\ 
 $P$ [days] &  $1.977150 \pm 0.000029$ & $1.977175 \pm 0.000011$ & $1.977170 \pm 0.000007$ & $1.977171 \pm 0.000011$ \\ 
 $t_C$ [days] & $0.000194 \pm 0.000270$ & $0.000007 \pm 0.000096$ & $-0.000038 \pm 0.000062$ & $-0.000017 \pm 0.000097$ \\ 
 $u_{1+}$ &  $0.76 \pm 0.06$ & $0.61 \pm 0.04$ & $0.43 \pm 0.03$ & -- \\ 
 $u_{1-}$ &  $0.32 \pm 0.07$ & $0.13 \pm 0.06$ & $0.02 \pm 0.07$ & -- \\ 
$\sigma_{r}$ [100 ppm] & $913.641 \pm 35.007$ & $425.538 \pm 10.116$ & $304.506 \pm 5.217$ & -- \\ 
$\sigma_{w}$ [100 ppm] & $21.297 \pm 0.097$ & $7.873 \pm 0.035$ & $4.626 \pm 0.021$ & --  \\ 
\hline
\\[\dimexpr-\normalbaselineskip+3pt]
\multicolumn{5}{c}{\textit{WASP-156 b}} \\ [\dimexpr-\normalbaselineskip+12pt]
\hline
$a / R_S$ &  $12.91 \pm 0.20$ & $12.92 \pm 0.28$ & $12.78 \pm 0.29$ & $12.88 \pm 0.15$ \\ 
 $R_P / R_S$ &  $0.06637 \pm 0.00060$ & $0.06729 \pm 0.00057$ & $0.06689 \pm 0.00050$ & $0.06687 \pm 0.00032$ \\ 
 $b$ &  $0.112 \pm 0.104$ &  $0.173 \pm 0.121$ & $0.198 \pm 0.127$ & $0.155 \pm 0.068$ \\ 
 $P$ [days] &  $3.836167 \pm 0.000085$ & $3.836304 \pm 0.000052$ & $3.836159 \pm 0.000048$ & $3.836218 \pm 0.000037$ \\ 
 $t_C$ [days] & $-0.000346 \pm 0.000350$ & $-0.000651 \pm 0.000214$ & $0.000121 \pm 0.000204$ & $-0.000261 \pm 0.000154$ \\ 
 $u_{1+}$ &  $0.80 \pm 0.05$ & $0.61 \pm 0.04$ & $0.53 \pm 0.03$ & -- \\ 
 $u_{1-}$ &  $0.76 \pm 0.06$ & $0.41 \pm 0.06$ & $0.29 \pm 0.06$ & -- \\ 
$\sigma_{r}$ [100 ppm] & $22.217 \pm 25.088$ & $249.831 \pm 6.241$ & $251.481 \pm 5.159$ & -- \\ 
$\sigma_{w}$ [100 ppm] & $12.994 \pm 0.034$ & $7.445 \pm 0.019$ & $6.302 \pm 0.017$ & --  \\ 
\hline
\end{tabular}
\end{table*}

%% file: table_plato.tex
\begin{table*}
\caption{Transit parameters of the giraffes from the two considered cases of possible observations: using all four camera groups or using just one.}
\label{tab:giraffes-plato}
\centering
\begin{tabular}{l c c c}
\hline
\hline
Parameter & LTT 9779 b& TOI-674 b & WASP-156 b\\
\hline
\\[\dimexpr-\normalbaselineskip+3pt]
\multicolumn{4}{c}{\textit{24 cameras}} \\ [\dimexpr-\normalbaselineskip+12pt]
\hline
$a / R_S$ &  $3.786 \pm 0.038$ & $12.00 \pm 0.34$ & $12.75 \pm 0.34$ \\ 
 $R_P / R_S$ &  $0.04564 \pm 0.00050$ & $0.11435 \pm 0.00103$ & $0.06584 \pm 0.00071$  \\ 
 $b$ &  $0.9233 \pm 0.0025$ &  $0.682 \pm 0.025$ & $0.21 \pm 0.14$ \\ 
 $P$ [days] &  $0.7920650 \pm 0.0000047$ & $1.977156 \pm 0.000027$ & $3.836109 \pm 0.000046$ \\ 
 $t_C$ [days] & $0.000006 \pm 0.000097$ & $0.000035 \pm 0.000200$ & $3.83627 \pm 0.00019$ \\ 
$u_{+}$ & $0.64 \pm 0.04$ & $0.63 \pm 0.05$ & $0.76 \pm 0.03$ \\ 
$u_{-}$ & $0.38 \pm 0.07$ & $0.00 \pm 0.07$ & $0.56 \pm 0.06$ \\ 
$\sigma_{r}$ [100 ppm] & $111.154 \pm 1.426$ & $293.602 \pm 9.936$ & $210.455 \pm 3.796$ \\ 
$\sigma_{w}$ [100 ppm] & $2.108 \pm 0.006$ & $22.834 \pm 0.067$ & $5.549 \pm 0.016$ \\ 
\hline
\\[\dimexpr-\normalbaselineskip+3pt]
\multicolumn{4}{c}{\textit{6 cameras}} \\ [\dimexpr-\normalbaselineskip+12pt]
\hline
$a / R_S$ &  $3.795 \pm 0.045$ & $11.92 \pm 0.34$ & $12.27 \pm 0.53$ \\ 
 $R_P / R_S$ &  $0.04553 \pm 0.00072$ & $0.11486 \pm 0.00139$ & $0.06783 \pm 0.00094$  \\ 
 $b$ &  $0.9228 \pm 0.0030$ &  $0.685 \pm 0.025$ & $0.35 \pm 0.12$ \\ 
 $P$ [days] &  $0.7920378 \pm 0.0000084$ & $1.977131 \pm 0.000046$ & $3.836087 \pm 0.000076$ \\ 
 $t_C$ [days] & $0.00056 \pm 0.00018$ & $0.00024 \pm 0.00039$ & $3.83632 \pm 0.00032$ \\ 
$u_{+}$ & $0.65 \pm 0.06$ & $0.67 \pm 0.06$ & $0.63 \pm 0.04$ \\ 
$u_{-}$ & $0.38 \pm 0.07$ & $0.10 \pm 0.08$ & $0.49 \pm 0.06$ \\ 
$\sigma_{r}$ [100 ppm] & $89.840 \pm 2.587$ & $178.388 \pm 19.262$ & $236.760 \pm 5.087$ \\ 
$\sigma_{w}$ [100 ppm] & $4.231 \pm 0.012$ & $44.872 \pm 0.131$ & $10.919 \pm 0.027$ \\ 
\hline
\end{tabular}
\end{table*}